\documentclass{iopjournal}

\usepackage{amsmath}     % Often needed for math symbols/environments

\usepackage{amssymb,graphicx,tikz,xcolor,colortbl,booktabs,multirow,adjustbox}

\newtheorem{theorem}{Theorem}

\newtheorem{proposition}{Proposition}
\newtheorem{definition}{Definition}

\usepackage[justification=justified,singlelinecheck=false]{subcaption}
\usepackage[utf8]{inputenc}
\usepackage[T1]{fontenc}
\usepackage{booktabs,makecell,graphicx}    % Better caption handling
\usepackage[abs]{overpic}
\usepackage[T1]{fontenc}
\usepackage[utf8]{inputenc}

\usepackage{graphicx,dcolumn,bm,hyperref}% add hypertext capabilities
\usepackage[mathlines]{lineno}% Enable numbering of text and display math

\usepackage{ragged2e} % Required for better justification
\usepackage{caption}

\usepackage{float}
\usepackage{algorithm, algpseudocode}
\usepackage{algorithmicx}

\DeclareCaptionJustification{justified}{\justifying} % Force 'justified' to use ragged2e
\def\bea{\begin{eqnarray}}
\def\eea{\end{eqnarray}}

\makeatletter
\g@addto@macro\appendix{%
  \setcounter{section}{0}%
  \setcounter{equation}{0}%
  \@addtoreset{equation}{section}% Automatically resets equation count per appendix section
}
\makeatother

\begin{document}

\articletype{Paper} %	 e.g. Paper, Letter, Topical Review...

\title{Adaptive Qubit Freezing Enables Robust Graph Partitioning for Divide-and-Conquer QAOA}

\author{Sokea Sang$^1$\orcid{0009-0006-0345-5870}, Leanghok Hour$^1$\orcid{0009-0001-5670-7481}, Dongmin Kim$^1$\orcid{0000-0003-0320-9865} and Youngsun Han$^{1,*}$\orcid{0000-0001-7712-2514}}

\affil{$^1$Department of AI Convergence, Pukyong National University, Busan 48513, Republic of Korea}

% \affil{$^2$Department, Institution, City, Country}

% \affil{$^*$Author to whom any correspondence should be addressed.}

\email{youngsun@pknu.ac.kr}

\keywords{Quantum Approximate Optimization Algorithm (QAOA), Divide-and-Conquer Quantum Computing, Graph Partitioning, Qubit Freezing, MaxCut, NISQ Computing}

\begin{abstract}
Separator-based divide-and-conquer QAOA can address graphs wider than the qubit capacity of one circuit, meaning the number of physical qubits usable by that circuit, but the Large Graph Partitioner (LGP) in DC-QAOA aborts when no small vertex separator exists. We introduce Frozen Large Graph Partitioning (FrozenLGP), a three-phase extension of this specific pipeline. It first runs standard LGP (Phase~1); after failure, it freezes a minimum node cut of each selected residual graph, conditional on the first budget-feasible separator (Phase~2); and it uses a deterministic width-feasible split as Coverage-Preserving Partitioning (CPP) for a final fallback (Phase~3). 
For a fixed frozen spin assignment, removed Ising couplings are folded exactly into local fields; assignment selection, QAOA, truncated reconstruction, and fallback ranking remain approximate. 
On the pooled high-connectivity partition benchmarks ($n \leq 1000$), the scalable driver achieves 89.8\% separator-based coverage without the CPP fallback at freeze budget $B_f = 3$, compared with 4.6\% for standard LGP alone.
In the 100-instance end-to-end benchmark, FrozenLGP's Phase~2 increases coverage without CPP from 76 to 91 instances in the selected best-restart outcomes. The deterministic CPP fallback handles the remaining nine.
CPP-inclusive coverage is 100\%, including all tested partitioning instances up to $10^4$ vertices. This is a termination guarantee, not a measure of
freezing-based recovery. When Phase~1 succeeds, both methods follow identical seeded circuit paths and attain identical approximation ratios. Compared with QAOA-in-QAOA, FrozenLGP achieves higher mean approximation ratios at each measured size, at the cost of more invoked subcircuit solves and greater aggregate quantum work. Under two simulated noise channels, the coverage extension persists while the reported output distributions of every method degrade under the stronger channel. In simulation, these results support FrozenLGP as a topology-robust front end for separator-based divide-and-conquer QAOA.
\end{abstract}

\section{Introduction}\label{sec:introduction}

% --- Fig. Environment ---
\begin{figure*}[htbp]
    \centering
    % Subfigure A
    \begin{subfigure}[b]{0.36\textwidth}
        \centering
        \includegraphics[width=\textwidth]{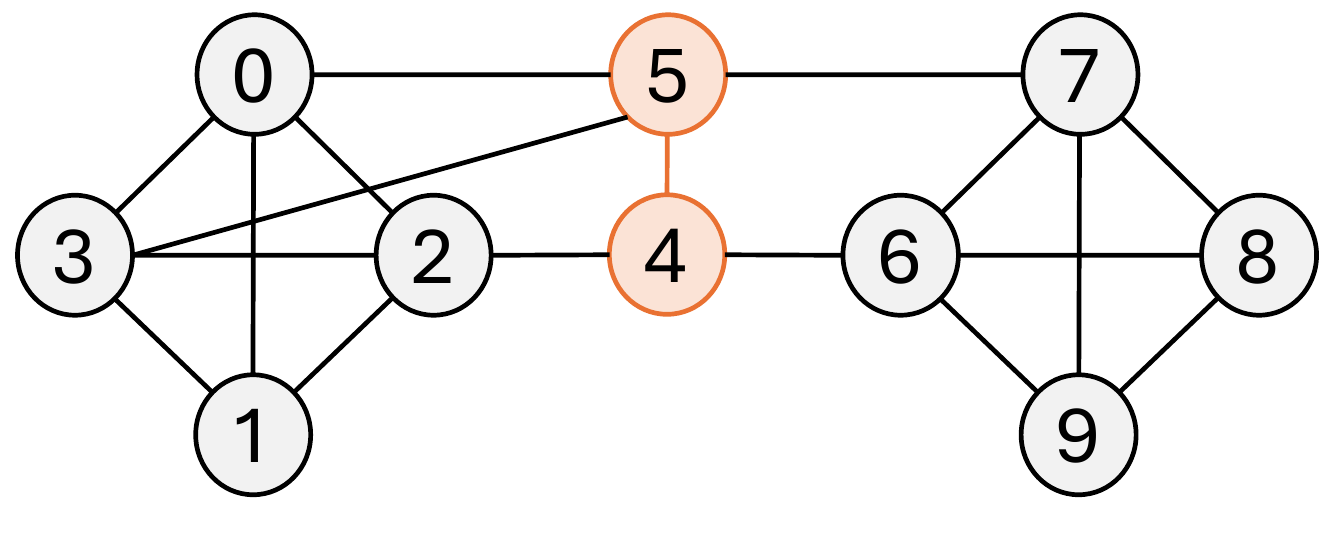}
        \caption{Sparse graph candidate.}
        \label{fig:1a}
    \end{subfigure}
    \hfill
    % Subfigure B
    \begin{subfigure}[b]{0.40\textwidth}
        \centering
        \includegraphics[width=\textwidth]{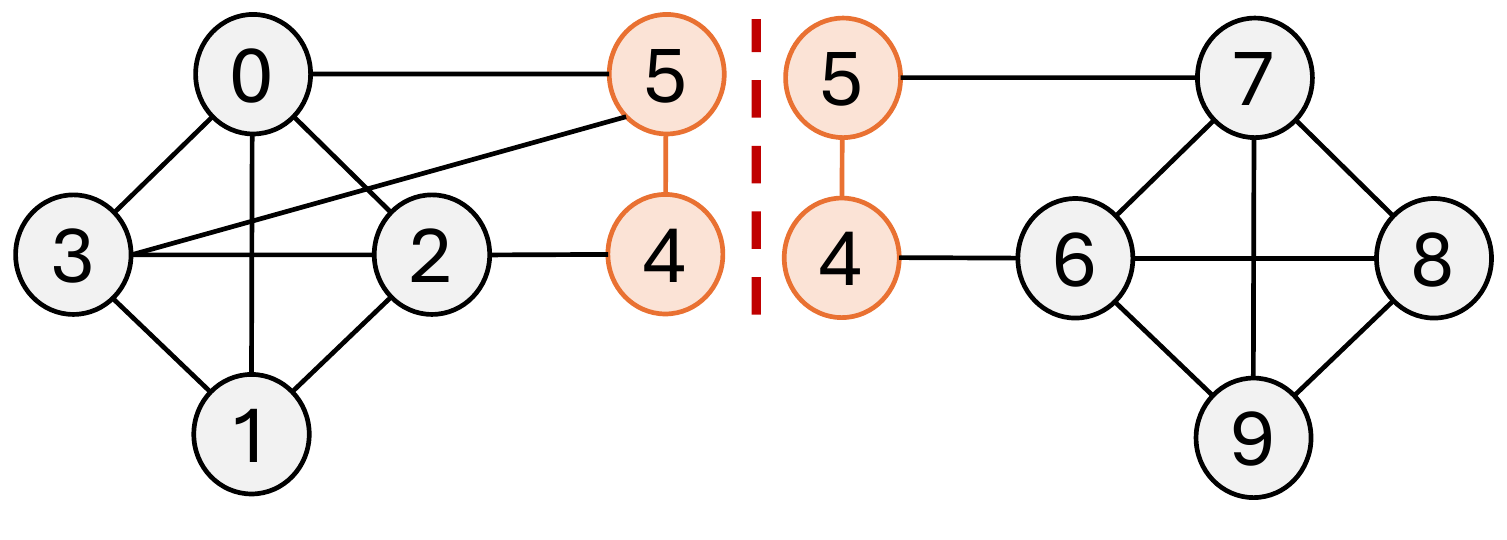}
        \caption{Valid LGP partition.}
        \label{fig:1b}
    \end{subfigure}
    % \hfill
    \vspace{0.5cm} % Add some vertical space between rows
    
    % Subfigure C
    \begin{subfigure}[b]{0.31\textwidth}
        \centering
        \includegraphics[width=\textwidth]{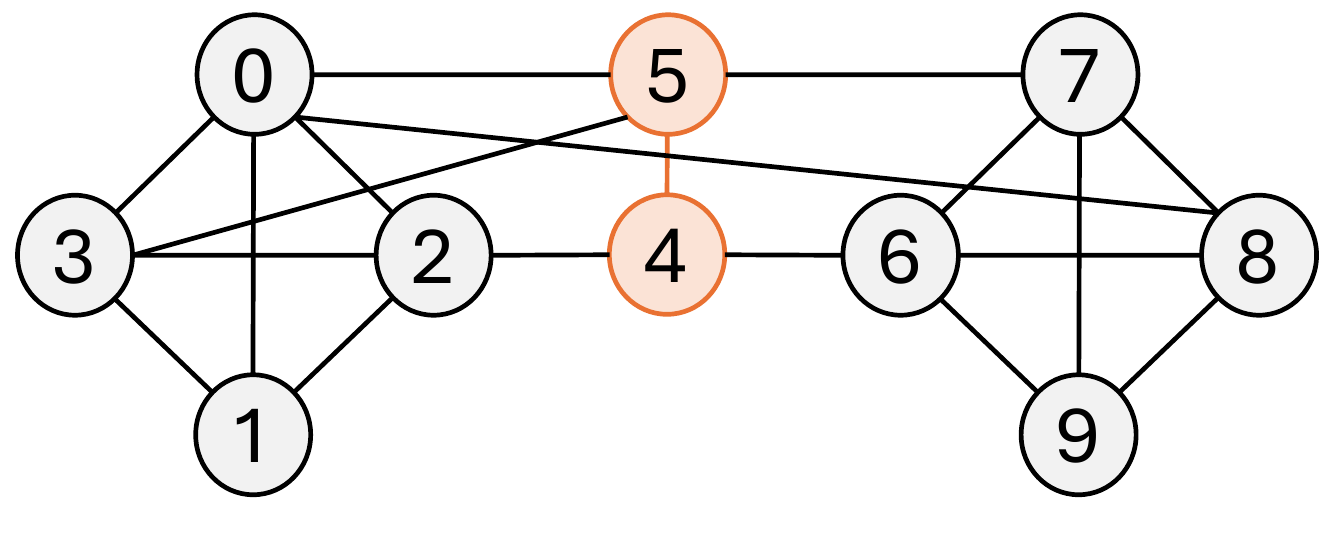}
        \caption{Dense graph (LGP fails).}
        \label{fig:1c}
    \end{subfigure}
    \hfill
    % Subfigure D
    \begin{subfigure}[b]{0.31\textwidth}
        \centering
        \includegraphics[width=\textwidth]{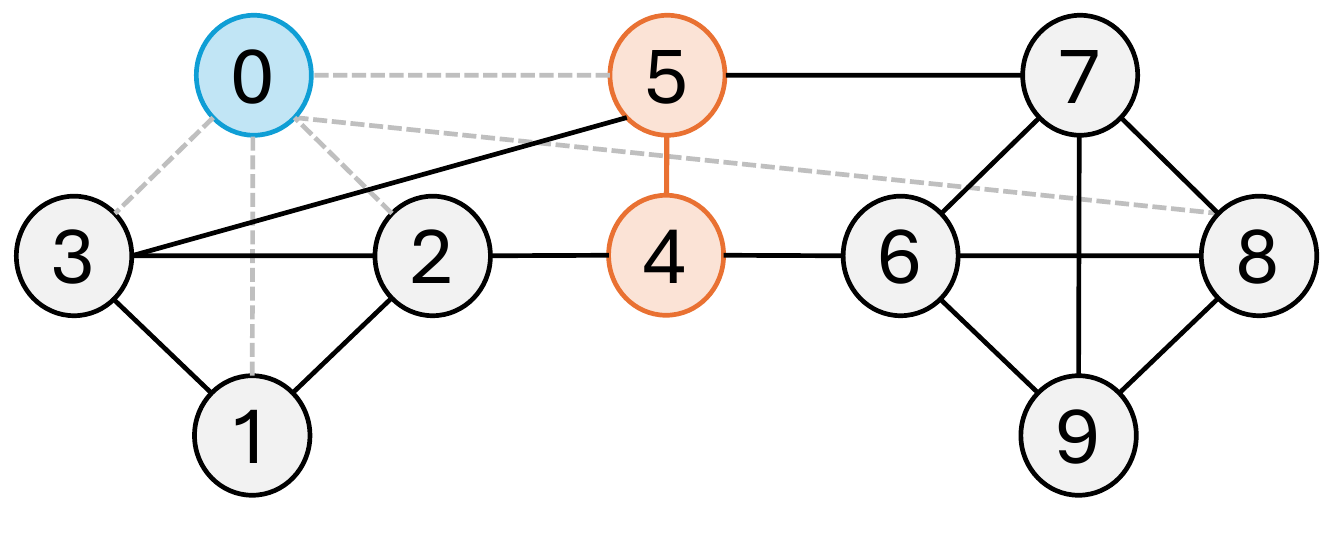}
        \caption{Node freezing (FrozenLGP).}
        \label{fig:1d}
    \end{subfigure}
    \hfill
    % Subfigure E
    \begin{subfigure}[b]{0.34\textwidth}
        \centering
        \includegraphics[width=\textwidth]{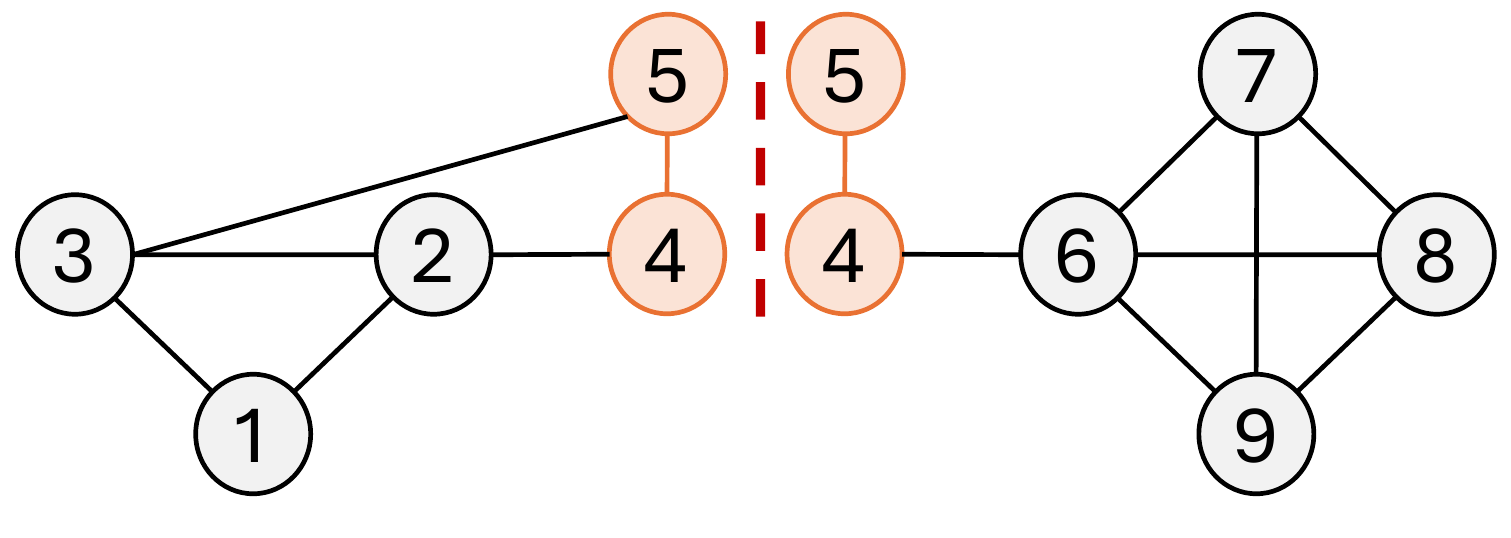}
        \caption{Recovered valid partition.}
        \label{fig:1e}
    \end{subfigure}
    
    \caption{Overview of the FrozenLGP partitioning mechanism. (a) A graph
    exceeding the NISQ qubit budget, with nodes 4 and 5 highlighted as a
    separator candidate. (b) Removing that pair bipartitions the sparse graph;
    the shared separator is duplicated into both sub-circuits and re-merged by
    MDR. (c) Additional crossing edges invalidate the same candidate in the
    denser schematic graph. (d) Conditioning node 0 removes its incident edges
    from the active topology and folds their couplings into local fields.
    (e) The highlighted candidate then separates the active graph. Panels
    (c)--(e) illustrate this candidate-level mechanism for readability; they do
    not assert that the exhaustive Phase~1 sweep fails on this exact ten-node
    topology. In the implemented pipeline Phase~2 starts only after every
    Phase~1 candidate of size at most $k-1$ has failed.}
    \label{fig:frozenlgp_overview}
\end{figure*}

The advent of Noisy Intermediate-Scale Quantum (NISQ)~\cite{nisq} devices has spurred the development of hybrid quantum-classical algorithms, most notably the Quantum Approximate Optimization Algorithm (QAOA)~\cite{qaoa_origin}, designed to tackle combinatorial optimization problems such as MaxCut~\cite{qaoa_bound_numerical_max_cut, fixed_angle_regular_max_cut, qaoa_transfer_param_max_cut, qaoa_max_cut_fermion}. In practice, QAOA is confined by the width of the qubit register and by the noise accumulated as circuits deepen, and distributed and hierarchical execution schemes have emerged as the main route around both limits~\cite{towards_distributed_qc, cutQC, quantum_cir_on_small_qpu, qarma}. All of these schemes share a common anatomy. A decomposition stage splits the problem graph into subgraphs small enough for the register; a quantum stage solves each subgraph; a reconstruction stage reassembles the pieces into a global solution.

The research effort behind this paradigm has been distributed unevenly across that anatomy. Reconstruction policies, merging schemes, and parallel execution have each received dedicated treatment in the original Divide-and-Conquer QAOA (DC-QAOA)~\cite{dc-qaoa} and in the extensions that followed, including QAOA-in-QAOA~\cite{qaoa-in-qaoa}, coupling-based QAOA~\cite{coupling-qaoa}, and parallel QAOA frameworks~\cite{paraqaoa}. These decomposition methods do not all share the same prerequisite. In separator-based DC-QAOA~\cite{dc-qaoa}, the decomposition stage presumes that a valid register-sized decomposition of the input can be obtained, and prior work does not address what happens when it cannot. Yet decomposition is the gatekeeper of the DC-QAOA workflow: it runs first, and when it fails, no downstream sophistication in solving or reconstruction is reached. We treat robust decomposition as a problem in its own right and supply the layer that this separator-based stack has been missing.

The empirical claims of this paper are restricted to unweighted MaxCut. The
partition operations are objective-independent at the graph level, and
fixed-spin substitution extends algebraically to pairwise Ising couplings, but
we do not claim demonstrated solution quality for weighted MaxCut or broader
Ising/QUBO objectives. Accordingly, the graph-problem framing below describes
structural applicability rather than validated performance beyond unweighted
MaxCut.

DC-QAOA~\cite{dc-qaoa} makes the gap concrete. Its Large Graph Partitioning (LGP) policy must identify a node separator whose removal disconnects the graph into exactly two subcomponents with no crossing edges (Fig.~\ref{fig:frozenlgp_overview}(a-b)); the Measurement Distribution Reconstruction (MDR) policy then reassembles the sub-solutions across that shared separator. The prerequisite breaks down on dense or highly connected networks, including dense Erd\H{o}s--R\'enyi (ER) graphs~\cite{random_ER_graph}, high-degree Barab\'asi--Albert (BA) graphs~\cite{BA_graph}, and complete graphs ($K_n$), where removing a small set of shared nodes rarely severs the graph. When LGP fails, the entire pipeline aborts and returns no
decomposition. In our partition-level benchmark the effect is stark: on the pooled high-connectivity families the baseline decomposes only 4.6\% of instances (Sec.~\ref{subsec:partition-scalability}). This motivates extending
LGP while retaining its separator-compatible path whenever that path exists.

To intuitively illustrate the motivation and mechanics of the FrozenLGP algorithm, Fig. \ref{fig:frozenlgp_overview} presents a comparative visual progression of graph partitioning under varying connectivity conditions. In the standard Divide-and-Conquer QAOA framework~\cite{dc-qaoa}, a large graph that exceeds the available hardware qubit budget $k$ (Fig. \ref{fig:frozenlgp_overview}(a)) must be divided. The LGP attempts to identify a separator path (e.g., nodes 4 and 5) whose removal completely disconnects the graph into exactly two independent subcomponents (Fig. \ref{fig:frozenlgp_overview}(b)). Because these subgraphs share only the separator nodes, their respective quantum states can be measured independently and recombined classically without losing boundary edge information.

However, this vertex-removal strategy is highly sensitive to graph density. When applied to a higher-connectivity graph (Fig. \ref{fig:1c}), additional crossing edges bridge the network in ways that bypass the proposed separator path. Consequently, removing nodes 4 and 5 no longer yields two disjoint subgraphs; that candidate remains invalid. The figure is a readable schematic of the candidate-level mechanism, not a claim that the exhaustive LGP sweep fails on this particular ten-node topology.

FrozenLGP resolves the general failure mode by integrating classical qubit freezing into the partitioning phase itself. Only after the complete Phase~1 sweep fails does the algorithm search for a fixed-residual minimum node cut. In the schematic trace, node 0 is conditioned (Fig. \ref{fig:1d}). By classically fixing its state to $+1$ or $-1$, its incident edges are removed from the active quantum topology and their energetic contributions are folded into linear Ising fields on neighbouring active qubits. The displayed separator path (nodes 4 and 5) then partitions the active graph (Fig. \ref{fig:1e}).
For graphs fundamentally incompatible with vertex-removal bipartitioning (such as complete graphs~\cite{dittmann2017menger, bondy1976graph}), we introduce Coverage-Preserving Partitioning (CPP), a deterministic fallback that prevents termination with a partition failure. CPP is introduced here as a termination
safeguard; it does not recover the separator-based route, and its use is reported separately from freezing-based recovery.

Robust decomposition can, of course, be bought in other ways, and design points clarify why we build the layer this way rather than adopting an existing route. QAOA-in-QAOA (QAOA$^2$)~\cite{qaoa-in-qaoa} partitions the graph \emph{arbitrarily} and solves a merging problem, so it does not require a vertex separator and does not fail on dense graphs; it instead pays for coverage with a cut-edge merging stage whose fidelity degrades as the discarded cross-partition weight grows. FrozenLGP makes the opposite trade in Phases~1 and 2: it preserves the shared separator, so boundary couplings stay inside the child Hamiltonians, and absorbs obstructing vertices as conditional local fields. Because QAOA$^2$ already achieves full coverage, the meaningful question is quality, and we benchmark the two head-to-head at scale in Sec.~\ref{subsec:qaoa2}, where FrozenLGP attains a higher mean approximation ratio at every measured size, at the cost of more invoked subcircuit solves, because separator preservation avoids the cut-edge loss that QAOA$^2$'s arbitrary partition incurs at every merge.

Our answer is the Frozen Large Graph Partition (FrozenLGP) algorithm, an adaptive decomposition framework that makes partitionability a property the pipeline can enforce rather than an assumption it must inherit. The mechanism is qubit freezing, applied at the partitioning layer. For each selected separator candidate $S$, FrozenLGP computes a minimum vertex cut of the fixed residual $G\setminus S$ via max-flow~\cite{max-flow-theory, max-min-flow_cut_theorem} and accepts the first budget-feasible pair; it does not minimize the frozen set jointly over every $S$. It classically fixes the returned vertices to spins of $+1$ and $-1$ so that the residual active graph bipartitions. Their eliminated interactions are folded into linear Ising fields on neighbouring active qubits, leaving each conditional sub-circuit Hamiltonian well-defined.

This work makes three contributions:
\begin{enumerate}
    \item \textbf{A budget-aware decomposition algorithm:} For each fixed
    residual $G\setminus S$, FrozenLGP computes a minimum-cardinality vertex
    cut via max-flow and accepts the first budget-feasible pair in deterministic
    order. This conditional minimality bounds the assignment count for that
    residual; it neither minimizes induced field magnitude nor certifies a
    global minimum over $S$. The sampled random-regular benchmarks exhibit an
    empirical recovery staircase consistent with the connectivity surplus. At $B_f = 3$, coverage without CPP reaches 89.8\% on the pooled high-connectivity partition benchmarks, versus 4.6\% for Phase 1 alone; in the 100-instance end-to-end benchmark, Phase 2 increases this separator-based coverage from 76 to 91 selected best-restart outcomes.
    \item \textbf{A deterministic termination safeguard:} When the separator-compatible phases fail, CPP supplies a width-feasible split so that the configured pipeline can return a candidate. It records omitted cross-partition edges for full-parent-objective ranking of the retained MDR candidates. The resulting CPP-inclusive 100\% coverage is a termination guarantee, reported separately from the freezing-based recovery in contribution 1; it does not guarantee preservation of the omitted interactions or approximation quality.
    \item \textbf{Width and per-circuit resource reduction:} By removing
    frozen vertices and their active two-body interactions, FrozenLGP reduces
    leaf width and the maximum two-qubit-gate count per circuit. Aggregate
    execution work is reported separately and can exceed that of alternative
    decompositions.
\end{enumerate}

The remainder of this manuscript is organized as follows. Section~\ref{sec:background} establishes the preliminaries of the MaxCut problem and the principles of qubit freezing. Section~\ref{sec:method} formalizes the proposed FrozenLGP method. Section~\ref{sec:evaluation} presents a comprehensive empirical evaluation of our approach across varying graph topologies. Section~\ref{sec:discussion} discusses the limitations and directions for future optimization. Finally, Section~\ref{sec:conclusion} concludes the paper.

\section{Background And Related Work}\label{sec:background}
~\ref{sec:dcqaoa} outlines the existing Divide-and-Conquer QAOA (DC-QAOA) framework~\cite{dc-qaoa}, specifically the Large Graph Partitioner (LGP) and Measurement Distribution Reconstruction (MDR) policies whose structural vulnerabilities on dense networks motivate the core contributions of this work.

To contextualize the proposed FrozenLGP algorithm, this section establishes the theoretical foundations and related methodologies that underpin our approach. 
We begin in Section~\ref{sec:prelim_qaoa_mapping} by reviewing the mathematical formulation of the MaxCut problem and its native mapping to QAOA via the generalized Ising model. Next, in Section~\ref{sec:prelim_qubit_freezing}, we introduce qubit freezing and bias folding. The folding identity preserves the Hamiltonian contribution of a frozen edge \emph{conditional on a specified classical spin assignment}; choosing and sampling those assignments remains part of the approximation pipeline.

% --- Fig. Environment ---
\begin{figure*}[htbp]
    \centering
    \includegraphics[width=\textwidth]{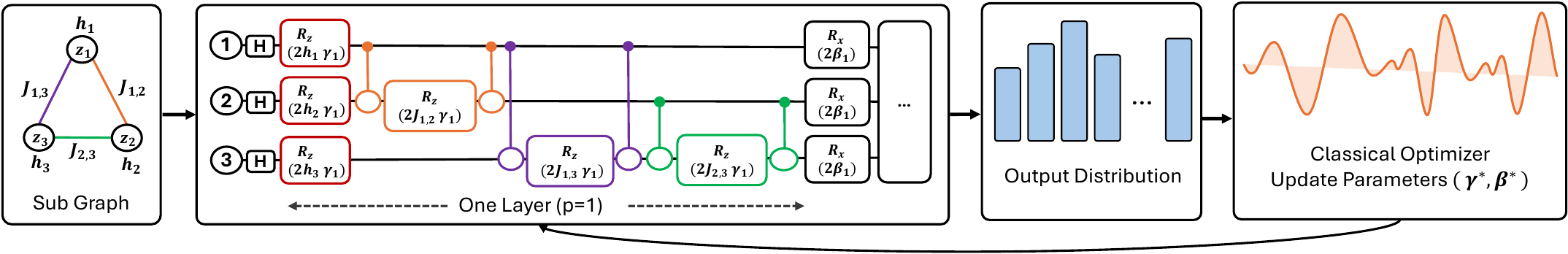}
    \caption{Schematic of a 1-layer ($p=1$) QAOA circuit with linear bias terms. With the Qiskit convention $R_P(\theta)=\exp(-i\theta P/2)$, the problem unitary uses local rotations $R_z(2 h_i \gamma_1)$ and CNOT--$R_z(2J_{i,j}\gamma_1)$--CNOT sequences for the $Z_iZ_j$ terms. The mixer uses $R_x(2\beta_1)$. The measured distribution is supplied to a classical optimizer that updates $(\gamma,\beta)$ to minimize the Ising Hamiltonian.}
    \label{fig:qaoa_circuit}
\end{figure*}

\subsection{Mapping MaxCut to QAOA via the Ising Model}
\label{sec:prelim_qaoa_mapping}

Combinatorial optimization problems such as MaxCut can be formulated as Quadratic Unconstrained Binary Optimization (QUBO) problems and mapped to an Ising model~\cite{ising_model, 19quantum}. Given an undirected graph $G=(V,E)$ with non-negative edge weights $w_{i,j}$, MaxCut assigns a binary label $z_i\in\{0,1\}$ to every vertex and maximizes $C(\mathbf z)=\sum_{(i,j)\in E}w_{i,j}(z_i\oplus z_j)$. All experiments in this paper use the \emph{unweighted} case $w_{i,j}=1$; weighted notation is retained only to state the folding identity generally.

To solve this on a quantum computer, we map the binary decision variables to quantum spin variables $s_i \in \{+1, -1\}$, corresponding to the eigenvalues of the Pauli-$Z$ operator $Z_i$. The generalized Ising Hamiltonian describing the energy landscape of the system is formulated as:
\begin{equation}\label{eq:isingH}
H_P = \sum_{(i,j) \in E} J_{i,j} Z_i Z_j + \sum_{i \in V} h_i Z_i
\end{equation}
where $J_{i,j}$ represents the coupling strength between connected qubits $i$ and $j$, and $h_i$ represents the localized linear bias (or external magnetic field) applied to a specific qubit $i$. 

For standard MaxCut~\cite{qaoa_bound_numerical_max_cut, fixed_angle_regular_max_cut, qaoa_transfer_param_max_cut, qaoa_max_cut_fermion} without fixed nodes, $h_i=0$ and $J_{i,j}=w_{i,j}/2$. Up to the additive constant $\frac12\sum_{(i,j)}w_{i,j}$, maximizing $C$ is therefore equivalent to minimizing Eq.~\eqref{eq:isingH}: aligned spins incur positive energy and anti-aligned spins negative energy. FrozenLGP introduces non-zero $h_i$ only after conditioning on frozen spins.

For a single layer ($p=1$), $U_P(\gamma_1)=e^{-i\gamma_1H_P}$ encodes the objective. Under Qiskit's definition $R_P(\theta)=e^{-i\theta P/2}$, a field term is implemented as $R_z(2h_i\gamma_1)$, and a coupling term as CNOT--$R_z(2J_{i,j}\gamma_1)$--CNOT~\cite{nielsen2010quantum}. Similarly, $U_M(\beta_1)=e^{-i\beta_1\sum_iX_i}$ uses $R_x(2\beta_1)$ on each qubit. 
%The factors of two follow directly from the half-angle gate convention and were verified against the corresponding matrix exponentials in our unit tests.

Once the parameterized quantum circuit is executed, the resulting quantum state $|\psi(\gamma, \beta)\rangle$ is measured in the computational basis. This yields an output distribution of bitstrings. The expectation value of the cost Hamiltonian, $\langle H_P \rangle = \langle \psi(\gamma, \beta) | H_P | \psi(\gamma, \beta) \rangle$, is estimated from this sampled distribution. 

This expectation value is then fed into a classical optimizer loop. The classical optimizer evaluates the objective function and calculates updated variational parameters $(\gamma^*, \beta^*)$~\cite{qaoa_origin, variational_peruzzo2014, kandala2017hardware}. These updated parameters are loaded back into the quantum circuit for the next iteration. This closed hybrid quantum-classical loop continues until the parameters converge, ideally driving the circuit to output a distribution tightly clustered around the optimal MaxCut bitstring.

Fig.~\ref{fig:qaoa_circuit} illustrates this loop for a three-qubit subgraph at $p=1$. Linear biases are independent local $Z$ rotations and are zero in the original unweighted MaxCut Hamiltonian; after a spin is frozen they encode its conditional interaction with each surviving neighbour.

Following the local biases, the quadratic edge couplings ($J_{1,2},J_{2,3},J_{1,3}$) use CNOT--$R_z$--CNOT sequences, and the $R_x$ mixer is applied before measurement. All rotation labels in this discussion denote the doubled Qiskit gate angles given above.

% --- Fig.: Identifying the Cross Node ---
\begin{figure*}[htbp]
    \centering
    \includegraphics[width=\textwidth]{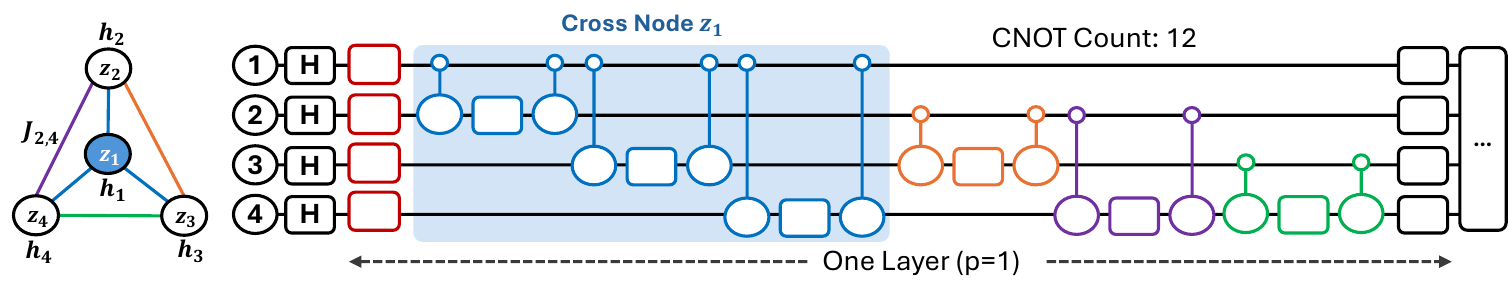}
    \caption{Standard QAOA circuit for a fully active 4-node network. Node $z_1$ acts as a central "cross node" connected to all other qubits ($z_2, z_3, z_4$). Mapping this fully quantum topology requires 12 CNOT gates for a single QAOA layer ($p=1$), with the blue-shaded region highlighting the extensive entangling overhead generated solely by $z_1$.}
    \label{fig:identify_cross_node}
\end{figure*}

\subsection{Principles of Qubit Freezing and Bias Folding}
\label{sec:prelim_qubit_freezing}

In the NISQ era, the fidelity of a quantum circuit is heavily constrained by its depth and the number of two-qubit entangling gates (e.g., CNOTs), which are significantly noisier than single-qubit rotations~\cite{qubit_mapping_nisq, swap_gates, hardware_aware_mapping, ibm_hardware_domonstrate}. Qubit freezing is a hybrid optimization technique originally developed to boost circuit fidelity by classically fixing the states of "hotspot" (high-degree) nodes, thereby stripping away the deep entangling sequences associated with them \cite{frozenQubits}.

Mathematically, freezing relies on the linearity of the Ising Hamiltonian. Consider an active quantum node $v$ connected to a high-degree node $u$. Their pairwise interaction in the cost Hamiltonian is given by $J_{u,v} Z_u Z_v$. If we decide to "freeze" node $u$, we remove it from the quantum circuit and treat it as a classical constant with a definitive spin state, $s_u \in \{+1, -1\}$. Consequently, the quantum operator $Z_u$ collapses into the scalar $s_u$. The two-body interaction term simplifies into a single-body linear term:
\begin{equation}\label{eq:freezefold}
J_{u,v} Z_u Z_v \xrightarrow{\text{freeze } u} (J_{u,v} s_u) Z_v
\end{equation}
This classical value is then folded into the existing linear bias (Ising field) of the neighboring quantum node $v$. The updated effective bias, $h_v'$, becomes:
\begin{equation}\label{eq:biasfold}
h_v' = h_v + J_{u,v} s_u
\end{equation}
For the fixed value $s_u$, this substitution is an algebraic identity: it preserves the contribution of edge $(u,v)$ in the \emph{conditional} Hamiltonian without a CNOT between $u$ and $v$. It does not by itself prove that a variational solver finds the global MaxCut optimum, nor that retaining only finitely many reconstructed samples is lossless.

% --- Fig.: Freezing the Node ---
\begin{figure*}[htbp]
    \centering
    \includegraphics[width=\textwidth]{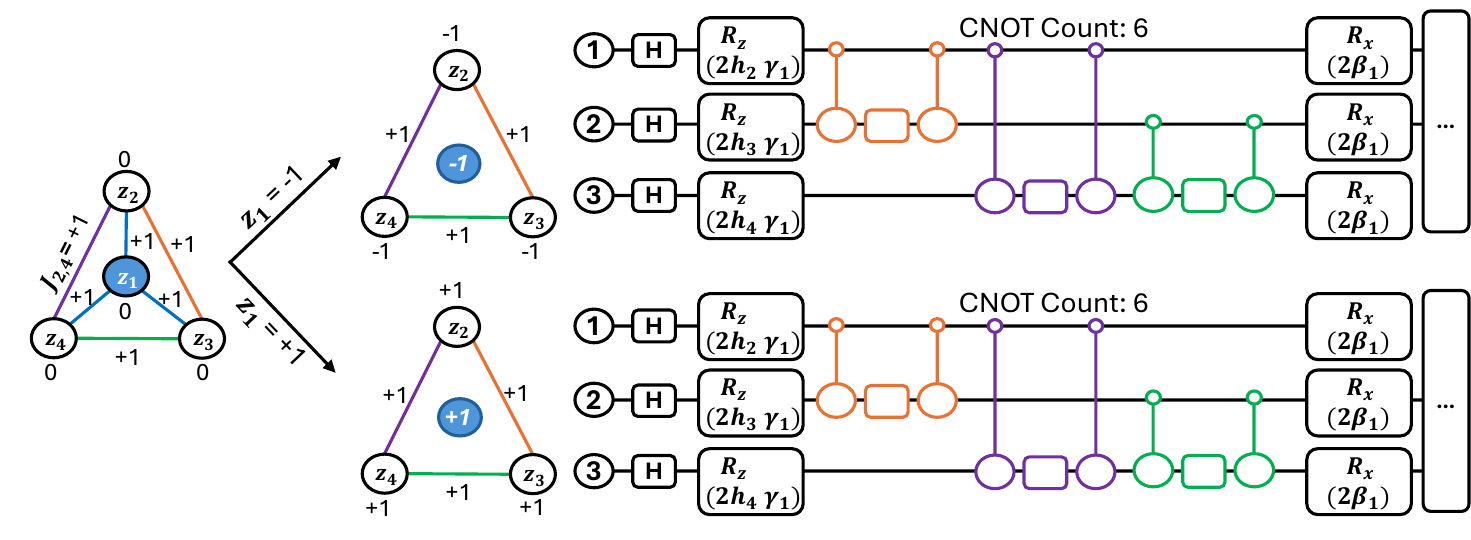}
    \caption{Circuit simplification via qubit freezing. Conditioning on $z_1=+1$ (left) or $z_1=-1$ (right) removes that node from the quantum circuit and reduces the illustrated CNOT count from 12 to 6. Its conditional coupling contributions are folded into $R_z(2h_i\gamma_1)$ rotations on the surviving qubits ($z_2,z_3,z_4$).}
    \label{fig:freeze_cross_node}
\end{figure*}

To illustrate this mechanism, consider the 4-node network depicted in Fig. \ref{fig:identify_cross_node}. The central node $z_1$ is heavily connected to the rest of the graph. If the entire network is mapped natively to a QAOA circuit, the high degree of $z_1$ generates a dense cascade of entangling operations, totaling 12 CNOT gates for just a single layer ($p=1$). The operations directly associated with $z_1$ (highlighted in the blue shaded region) dominate the circuit's error budget.

Fig. \ref{fig:freeze_cross_node} demonstrates the architectural transformation when $z_1$ is frozen. The node is physically removed from the quantum register, instantly eliminating the 6 CNOT gates associated with its edges. The target graph branches into two distinct classical sub-problems: one where $s_1 = +1$ and one where $s_1 = -1$. In both branches, the surviving qubits ($z_2, z_3, z_4$) absorb the frozen interaction via updated local biases. At the circuit level, these biases are injected natively by adjusting the rotation angles of the initial $R_z$ gates for each qubit.

Freezing one qubit creates two conditional branches. For any edge-only MaxCut Hamiltonian, weighted or unweighted, with zero linear fields, reversing every spin leaves the cut value unchanged, $C(\mathbf{s})=C(-\mathbf{s})$~\cite{qaoa_bound_numerical_max_cut, fixed_angle_regular_max_cut, qaoa_transfer_param_max_cut, qaoa_max_cut_fermion}. Paired frozen assignments can then share one quantum evaluation up to a global bit flip. Initial fields or biases inherited from an earlier recursion level break this symmetry; in that case the implementation enumerates both branches. \ref{sec:hbias} states the guard precisely.

While prior works~\cite{frozenQubits, do_qaoa} utilized freezing primarily as an error-mitigation strategy to reduce CNOT depth on sparse graphs, our FrozenLGP framework repurposes this mathematical property for structural graph recovery. By conditionally removing selected vertices from the active graph while retaining their incident couplings as local fields, the method changes residual connectivity so that LGP can recover some dense instances that would otherwise fail, a motivation we formalize in the following section.

\begin{figure*}[htbp]
    \centering
    \includegraphics[width=\textwidth]{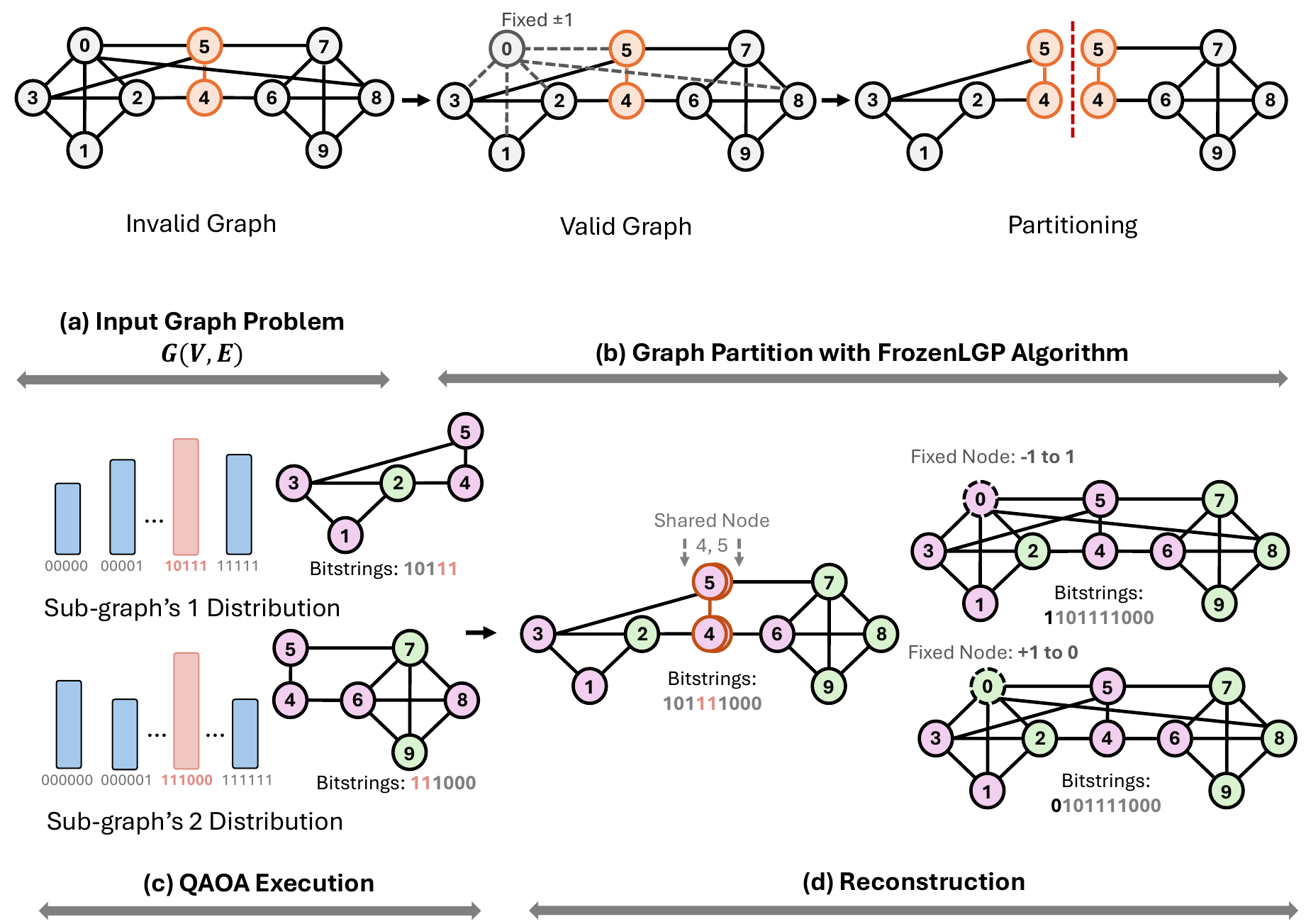}
    \caption{Schematic FrozenLGP workflow using the graph from
    Fig.~\ref{fig:1c}. (a)~The highlighted separator candidate, nodes 4 and 5,
    does not disconnect the input graph. (b)~Conditioning node 0 on each
    required spin assignment folds its incident couplings into local fields;
    nodes 4 and 5 then form the displayed shared separator and are duplicated
    in both child graphs. (c)~QAOA produces a measurement distribution for
    each conditional child graph, from which retained bitstrings are selected.
    (d)~Extended MDR merges child strings that agree on shared nodes 4 and 5
    and restores node 0 using $s_0=-1\mapsto 1$ and $s_0=+1\mapsto 0$.
    This readable ten-node drawing traces one separator candidate; it is not a
    certified exhaustive Phase~1-failure instance. The production algorithm
    invokes Phase~2 only after all Phase~1 candidates of size at most $k-1$
    have failed.}
    \label{fig:overall-workflow}
\end{figure*}

\section{Proposed Methodology}
\label{sec:method}

This section formalizes the decomposition layer. Frozen Large Graph Partitioning (FrozenLGP) augments the separator-dependent DC-QAOA pipeline with a recovery phase and a termination fallback phase. As established in Section~\ref{sec:prelim_qubit_freezing}, fixing a spin folds its incident couplings exactly into the conditional Hamiltonian of its active neighbours; the end-to-end solver remains approximate because assignments, QAOA samples, MDR truncation, and Phase~3 candidate ranking are algorithmic choices.

The partition logic is objective-independent at the graph level, and the
conditional substitution is valid for pairwise Ising terms. Nevertheless, all
end-to-end quality claims in this paper concern unweighted MaxCut. Inherited or
initial one-body fields disable the global spin-flip reduction and require the
full frozen-assignment enumeration described in \ref{sec:hbias}.

The central insight is to use that conditional substitution at the partitioning layer: selectively freezing obstructing vertices can expose a separator while retaining their assignment-dependent edge terms as local fields. Fig.~\ref{fig:overall-workflow} schematically follows the highlighted separator candidate through conditional freezing, two child-graph QAOA distributions, and heuristic MDR reconstruction with the frozen label restored.

\subsection{Overview of the FrozenLGP Pipeline}
\label{sec:pipeline}

Given an input graph $G = (V, E)$ with $|V| = n > k$, where $k$ denotes the hardware qubit budget per sub-circuit, the FrozenLGP pipeline executes three partition phases in strict priority order. Fig.~\ref{fig:overall-workflow}(a) and (b) schematically illustrate the key transition: the highlighted candidate nodes 4 and 5 do not disconnect the input, whereas conditioning node 0 exposes them as a shared separator before the two child graphs are handed to sub-circuit execution.

\emph{Phase 1 (Standard LGP).} The algorithm executes the LGP procedure of~\ref{dc_qaoa_LGP} directly. If a valid bipartition is identified, FrozenLGP returns that partition unchanged: no vertex is frozen and CPP is not invoked. The partition and subsequent seeded circuit path are therefore the same as in DC-QAOA implementation.

\emph{Phase 2 (Adaptive Qubit Freezing).} Phase~2 activates only after the complete Phase~1 sweep fails. It revisits candidate separators $S$ in increasing cardinality and deterministic node order. For each $S$ whose residual $G\setminus S$ remains connected, it computes a minimum node cut $F$ of that \emph{fixed residual graph}. The first pair $(S,F)$ satisfying the freeze budget and producing at least two residual components is returned. Thus $F$ is minimum conditional on the current $S$; the procedure does not minimize $|F|$ jointly over all separators.

\emph{Phase 3 (Coverage-Preserving Partitioning).} Phase~3 is a deterministic fallback (Sec.~\ref{sec:cpp}) that forms two vertex-balanced induced subgraphs and records edges omitted between their non-shared portions. The halves need not be connected; ``coverage-preserving'' means that the configured pipeline can continue and return a candidate instead of terminating with a partition failure. It does not mean that CPP preserves all graph edges or guarantees approximation quality.
In the LGP/MDR route retained by Phases 1 and 2, boundary edges remain in the child subproblems through the duplicated separator and MDR enforces agreement on the shared bits, whereas CPP omits edges crossing between the non-shared child portions from both child Hamiltonians.

The FrozenLGP orchestrator implements this control flow, invokes a QAOA
sub-circuit for the conditional bias-folded Hamiltonian
(\ref{sec:hbias}) for each required frozen-spin assignment, and retains
the assignment whose reconstructed candidates contain the highest evaluated
parent-graph cut. Extended MDR then returns at most the top $t$ parent
bitstrings (\ref{sec:mdr}). Strict phase priority ensures that freezing
and CPP are absent whenever Phase~1 succeeds; it does not imply equal total work
on instances that require later phases.

\subsection{Phase~2: Adaptive Qubit Freezing}
\label{sec:phase2}

After Phase~1 fails, Phase~2 performs a second exhaustive sweep over
$|S|=1,\ldots,k-1$. Node labels are sorted by type name and representation,
and combinations are traversed lexicographically, making candidate order
explicit. Disconnected or smaller-than-two-vertex residuals are skipped.
For each connected residual $G'=G\setminus S$, the production routine calls
\texttt{networkx.minimum\_node\_cut} without fixed terminals. In the pinned
NetworkX 3.6.1 environment~\cite{hagberg2008exploring} this global routine constructs a node-split
auxiliary digraph, selects a minimum-degree reference vertex, and solves a
sequence of local source--sink flow problems against its non-neighbours and
non-adjacent neighbour pairs. The returned $F$ has minimum cardinality among
sets that disconnect this particular $G'$~\cite{max-flow-theory,
max-min-flow_cut_theorem, graph_theory_west2001introduction}.

If $0<|F|\leq m_{\max}$ and $G\setminus(S\cup F)$ has at least two
components, the cut is feasible. Exactly two components are used directly;
if there are more, the largest component forms one node group and the union
of all remaining components forms the other. The latter group need not be
connected, but each induced group remains a valid Ising subproblem. Frozen
vertices are excluded from both quantum sub-circuits, while $S$ is shared.
Complete residual graphs are an explicit edge case: NetworkX returns a cut
of size $|V(G')|-1$, whose deletion leaves only one vertex, so the candidate
is rejected. If no ordered candidate is feasible, control passes to Phase~3.

Algorithm~\ref{alg:frozenlgp} summarizes the complete three-phase partitioning
procedure: Phase~1 retains standard LGP, Phase~2 applies the conditional
minimum-node-cut recovery described above, and Phase~3 invokes CPP only when
both separator-compatible phases fail. The
\textsc{GlobalMinNodeCut} call in Phase~2 is expanded in
Algorithm~\ref{alg:global-min-node-cut}, which mirrors the pinned NetworkX
3.6.1 routine used in the experiments.
\begin{algorithm}[t]
\caption{FrozenLGP: adaptive qubit-freezing partitioner.}
\label{alg:frozenlgp}
\begin{algorithmic}[1]
\Require Graph $G=(V,E)$; qubit budget $k$; freeze budget $m_{\max}$.
\Ensure  Sub-graphs $(G_1,G_2)$, frozen set $F$, and Phase~3 cut-edge list
         $E_{\mathrm{cut}}$.
\For{$s\gets 1$ \textbf{to} $k-1$}\Comment{Phase~1: standard LGP}
  \ForAll{$S\subseteq V$ with $|S|=s$}
    \If{$G\setminus S$ has exactly $2$ components $V_1',V_2'$}
       \State \Return $(G[V_1'\cup S],\, G[V_2'\cup S],\, \emptyset, \emptyset)$
    \EndIf
  \EndFor
\EndFor
\For{$s\gets 1$ \textbf{to} $k-1$}\Comment{Phase~2: MVC freeze fallback}
  \ForAll{$S\subseteq V$ with $|S|=s$}
    \If{$G\setminus S$ is connected and has at least two vertices}
      \State $F\gets\textsc{GlobalMinNodeCut}(G\setminus S)$
    \Else
      \State \textbf{continue}
    \EndIf
    \If{$0<|F|\leq m_{\max}$ \textbf{and}
        $G\setminus(S\cup F)$ has at least $2$ components}
      \State $(V_1',V_2')\gets\textsc{TwoGroups}(G\setminus(S\cup F))$
      \State \Return $(G[V_1'\cup S],\, G[V_2'\cup S],\, F,\,\emptyset)$
    \EndIf
  \EndFor
\EndFor
\State $(G_1,G_2,E_{\mathrm{cut}})\gets\textsc{CoveragePreservingPartition}(G)$
\State \Return $(G_1,G_2,\emptyset,E_{\mathrm{cut}})$\Comment{Phase~3}
\end{algorithmic}
\end{algorithm}

The fixed-residual guarantee is stated formally in
Proposition~\ref{prop:fixed-residual-mvc}. For a fixed connected residual
$G\setminus S$, no smaller freeze set can disconnect it. The implementation
then accepts the \emph{first} budget-feasible pair in its ordered $S$ sweep;
it does not compare feasible cuts across all $S$. A driver-agreement audit
reported in Table~\ref{tab:driver-agreement} of
\ref{app:driver-agreement} confirms the
distinction: the returned cardinality matches a global-over-$S$ reference in
51 of 69 comparable cases, with a maximum excess of four vertices. The
freeze budget bounds the assignment factor at one partition, but recursive
factors can multiply (\ref{sec:overhead}). For the $n\le24$
end-to-end studies we use the exhaustive routine above; the $n$ up to
$10^4$ preprocessing study uses the separately identified scalable driver
described in \ref{app:driver-agreement}. The separate recursive-work audit in
Table~\ref{tab:recursive-work} uses its recorded mixed policy: the exact driver
for subproblems of at most 20 vertices and the scalable driver above 20.

\subsection{Coverage-Preserving Partitioning Fallback}
\label{sec:cpp}

For graphs in which no separator of size $\leq k-1$ disconnects $G$ even
after freezing $m_{\max}$ vertices, both Phase~1 and Phase~2 fail by
construction. The canonical examples are the complete graphs $K_n$ (all-to-all graph)~\cite{dittmann2017menger, graph_theory_west2001introduction}, in
which every vertex-subset removal leaves the residual $K_{n-|S|}$
connected, and near-complete dense Erd\H{o}s--R\'enyi graphs~\cite{random_ER_graph} in which the
probability of finding a disconnecting separator vanishes rapidly with
edge density. To prevent the configured pipeline from
returning no candidate when both separator-compatible phases fail, Phase~3 invokes the Coverage-Preserving Partitioning (CPP) fallback. The name refers only to operational coverage: an instance is covered when the partition-and-reconstruction pipeline returns at least one candidate bitstring under the stated resource limit.

CPP departs from vertex-removal partitioning entirely. Given an input
graph $G=(V,E)$ with $n\geq 4$, it sorts labels by type name and
representation and splits the list at $\mu=\lfloor n/2\rfloor$.
The chunks $V_A=\{v_1,\dots,v_\mu\}$ and
$V_B=\{v_\mu,v_{\mu+1},\dots,v_n\}$ share precisely $v_\mu$,
which plays the role of a one-node separator in the subsequent
reconstruction step. Each chunk is then promoted to an \emph{induced}
sub-graph $G_A=G[V_A]$ and $G_B=G[V_B]$, while every edge $(u,v)\in E$
with $u\in V_A\setminus\{v_\mu\}$ and $v\in V_B\setminus\{v_\mu\}$ is
excluded from both induced sub-graphs and recorded as a \emph{cut edge}
together with its weight $w_{u,v}$. The two induced sub-graphs are then
handed to the standard FrozenLGP pipeline.

CPP does not restore the omitted edges to the quantum Hamiltonians. Instead,
after MDR has retained at most $t$ parent candidates, the production
rule evaluates each complete candidate directly on the original parent graph,
\begin{equation}
\label{eq:cppscore}
C_G(\mathbf{z})
\;=\;\sum_{(u,v)\in E} w_{u,v}\,\bigl(z_u\oplus z_v\bigr),
\end{equation}
and ranks candidates by $C_G(\mathbf z)$, using normalized MDR support only
as a deterministic sub-unit tie-breaker. This costs
$\mathcal{O}(|E|t)$ classical operations and no additional circuit
evaluation. It corrects the ranking of candidates already present in the
retained set; it cannot recover high-quality bitstrings discarded earlier by
top-$t$ truncation. Table~\ref{tab:cpp-ablation} in
\ref{app:cpp-ablation} compares this direct rule
with no rescoring and the multiplicative, additive, and normalized
weighted rules on paired circuit outputs.

Each CPP child contains at most $\lceil n/2\rceil$ vertices plus the one
shared boundary vertex, so repeated balanced splitting reduces width
geometrically. Connectivity is not promised. In the dense-family stress
benchmark of \ref{app:cpp-ablation}, CPP fired on eight of 40 graphs; those cases omitted on average 52.4\% of total edge weight and as much as 69.2\%. CPP is therefore a graceful-degradation path: it provides a deterministic termination guarantee, while separator-based recovery, approximation quality, and the retained candidate distribution are separate outcomes.

\section{Evaluation}\label{sec:evaluation}

Every solution-quality experiment in this section uses unweighted MaxCut with
$w_{ij}=1$. The $n\le10^4$ partitioning study evaluates graph structure and
preprocessing coverage only; it does not extend the quality claims to weighted
MaxCut or broader Ising/QUBO objectives.
\begin{table*}[t]
\caption{\label{tab:graph-families}%
Graph families used in the partition-level scalability evaluation. The
vertex connectivity $\kappa$ relative to the separator budget $k-1=5$ ($k = 6$ default)
determines the expected partitioning regime: Phase~1 suffices when
$\kappa \le k-1$; Phase~2 freezing is required when $\kappa > k-1$ (with
predicted budget threshold $B_f = \kappa-(k-1)$); CPP engages only where
local connectivity exceeds $k-1+B_f$. Sizes per scale category: small
$n\!=\!10$--$100$ (5 seeds), medium $n\!=\!150$--$500$ (3 seeds), large
$n\!=\!600$--$1000$ (3 seeds), extreme $n\!=\!2000$--$10{,}000$ (2
seeds).}
% \begin{ruledtabular}
\centering
\resizebox{\columnwidth}{!}{%
\begin{tabular}{llllll}
\hline
Family & Generator & Degree structure & $\kappa$ & Expected regime & Study \\
\hline
BA $d{=}3$ & Barab\'asi--Albert, $m{=}3$ & heavy-tailed, $\delta{=}3$ & ${\approx}3$ & Phase 1 & Sec.~\ref{subsec:partition-scalability} \\
BA $d{=}7$ & Barab\'asi--Albert, $m{=}7$ & heavy-tailed, $\delta{=}7$ & ${\approx}7$ & Phase 2 ($B_f{\ge}2$) & Sec.~\ref{subsec:partition-scalability} \\
Dense ER & $G(n,p)$, $p{=}12/(n{-}1)$, connected & Poisson, $\langle d\rangle{=}12$ & varies & mixed (Phases 1--3) & Sec.~\ref{subsec:partition-scalability} \\
Two-cluster BA & $2\times$BA($n/2$, $2$) $+$ 2 bridges & heavy-tailed, bridged & $2$ & Phase 1 (control) & Sec.~\ref{subsec:partition-scalability} \\
Random $d$-regular & pairing model, $d \in \{3,\dots,10\}$ & uniform, $\delta{=}d$ & $d$ (w.h.p.) & Phase 1 iff $d{\le}5$; else Phase 2 & Sec.~\ref{subsec:regular-scalability} \\
\hline
\end{tabular}
% \end{ruledtabular}
}
\end{table*}

The evaluation addresses four questions. First, how often does the
three-phase partitioner return width-feasible leaves when standard LGP fails?
Second, on instances for which both pipelines execute, does the strict phase
priority preserve the same circuits and reconstructed answers? Third, what
aggregate optimisation, shot, and two-qubit-gate--shot work is incurred by
recursive conditional solves, rather than only by one leaf circuit? Fourth,
how do the retained candidate distributions change under explicitly specified
simulated noise models? Partitioning coverage is an operational property of
the implemented pipeline rather than a statement about universal quantum
solvability, and
preprocessing time is reported separately from quantum workload. The scaling
experiments use the polynomial-time driver identified below; the controlled
end-to-end comparisons use the exact driver and exact ground truth. The
QAOA-in-QAOA study contains both an at-scale arm with aligned per-subproblem
settings and a smaller counter-level arm with shared objective-evaluation and
shot caps. Section~\ref{sec:discussion} interprets their distinct
quality--work trade-offs rather than treating either arm as evidence of
quantum advantage.

\subsection{Graph Partitioning Scalability}\label{subsec:partition-scalability}
We distinguish separator-based coverage without CPP (Phases~1 and 2) from CPP-inclusive operational coverage (all three phases). The former includes instances already solved by Phase~1 and those recovered by adaptive freezing, with no CPP invocation anywhere in their
decomposition. Thus “freezing-only coverage” below means coverage with CPP excluded, not the fraction newly recovered by Phase~2 alone. On the pooled high-connectivity
families at $B_f=3$, these coverage rates are 89.8\% without CPP and 100\%
including CPP, compared with 4.6\% for Phase~1 alone
(Table~\ref{tab:budget-sweep}). In the separate end-to-end benchmark, the
corresponding best-restart counts are 76 Phase-1 outcomes, 15 additional
Phase-2 outcomes, and nine CPP outcomes: $76\to91$ without CPP and 100 with
CPP (Sec.~\ref{subsec:e2e-quality}).

\begin{table}[!htbp]
\caption{\label{tab:budget-sweep}%
Freeze-budget sweep on the high-connectivity families (BA $d{=}7$ and
dense ER pooled, $n\le1000$). Coverage$^{\mathrm{F}}$ counts instances
decomposed by freezing alone (no CPP); FrozenLGP coverage including CPP
is 100\% throughout. $\rho$ is the graph reduction ratio
$|F_{\mathrm{total}}|/n$ averaged over successful runs.
NRL (\ref{dc_qaoa_NRL}, Eq.~\eqref{eq:nrl}) is reported over each
method's \emph{successful} runs, so the two populations differ: DC-QAOA's
$1.90$ is averaged over the $4.6\%$ of instances it can decompose, whereas
FrozenLGP's is averaged over all instances (the dense majority it additionally
recovers raises the duplication). On the common solvable set the strict
phase-priority guarantee makes the two NRLs \emph{identical} (FrozenLGP freezes
nothing there); the meaningful trend is the monotone decrease $2.95\!\to\!2.82$
as the freeze budget grows and removes vertices from the structural leaf graphs.
Runtime is the mean over all attempts, including attempts that terminate in a
reported partition failure.}
\centering
% \resizebox{\columnwidth}{!}{%
\begin{tabular}{lccccc}
\hline
Method & $B_f$ & Coverage$^{\mathrm{F}}$ & Runtime [s] & $\rho$ [\%] & NRL \\
\hline
DC-QAOA (Phase 1) & --- & 4.6\%  & 0.149 & 0 & 1.90 \\
FrozenLGP & 1  & 9.3\%  & 0.226  & 3.9  & 2.95 \\
FrozenLGP & 2  & 79.6\% & 0.011  & 33.5 & 2.90 \\
FrozenLGP & 3  & 89.8\% & 0.009  & 37.5 & 2.82 \\
FrozenLGP & 4  & 90.7\% & 0.009  & 37.8 & 2.82 \\
FrozenLGP & 5  & 90.7\% & 0.009  & 37.8 & 2.82 \\
FrozenLGP & 10 & 90.7\% & 0.008  & 37.8 & 2.82 \\
\hline
\end{tabular}
% }
\end{table}

\begin{figure}[t]
    \centering
    \includegraphics[width=0.8\columnwidth]{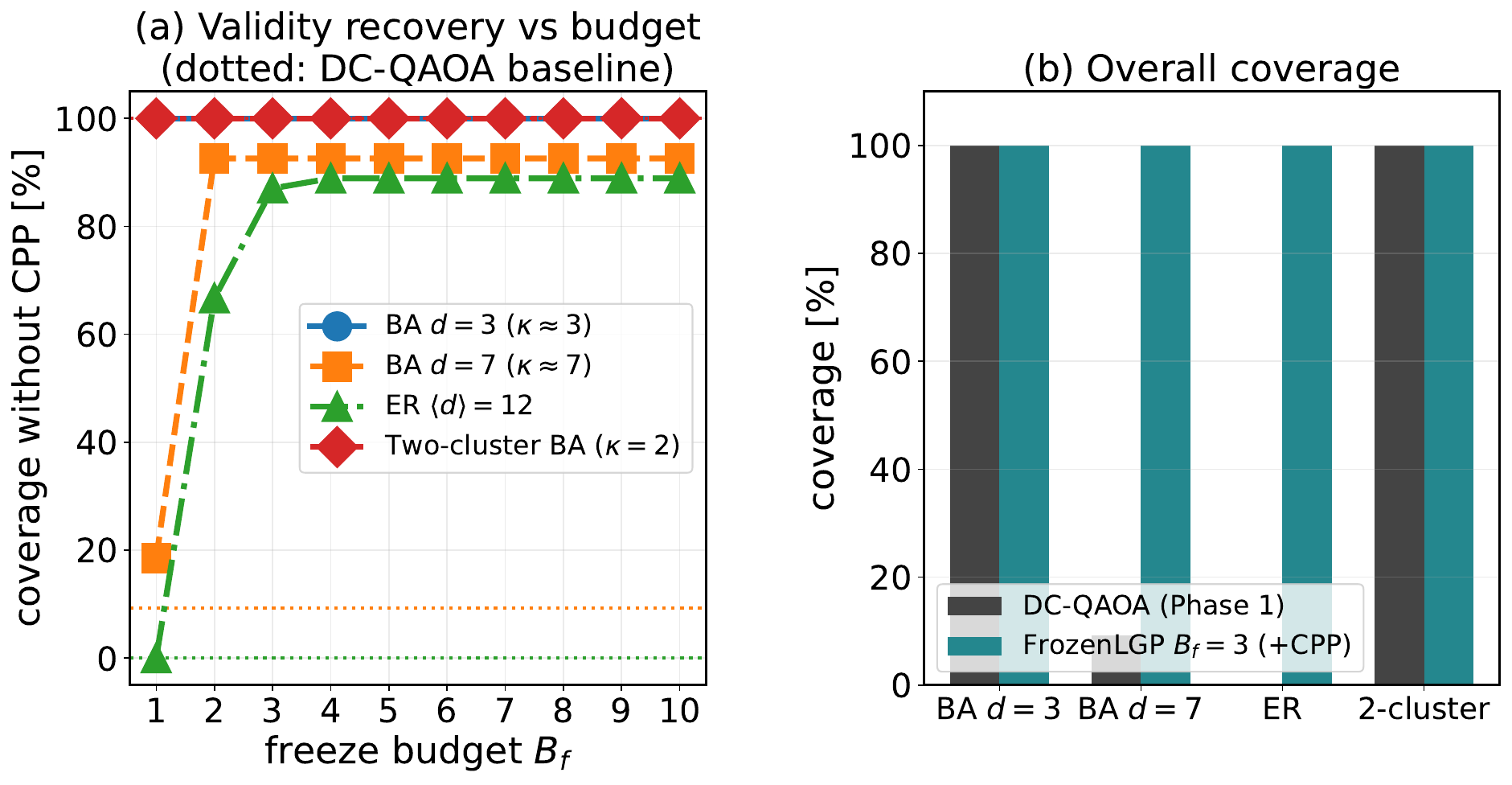}
    \caption{(a) Fraction of instances decomposed by freezing alone
    (CPP excluded) versus freeze budget $B_f$, per family ($n\le1000$
    pooled); dotted lines mark the Phase-1-only baseline. Recovery
    exhibits a sharp threshold at $B_f = \kappa-(k-1)$ and saturates by
    $B_f{=}4$. (b) Overall coverage including the CPP fallback:
    FrozenLGP reaches 100\% on every family, while the baseline fails on
    the high-connectivity families.}
    \label{fig:scal-coverage}
\end{figure}

\emph{Experimental setup.}
We benchmark the complete three-phase partitioning pipeline of
Sec.~\ref{sec:method} against the DC-QAOA baseline restricted to Phase~1
(standard vertex-separator search; freezing and CPP disabled), so that any
difference between the two methods is attributable solely to the adaptive
freezing mechanism and the CPP fallback. Both methods decompose each input
graph fully, i.e.\ recursively until every leaf sub-graph fits the qubit
budget $k=6$. Table~\ref{tab:graph-families} summarises all graph
families used in the partition-level evaluation. This subsection studies
the four heterogeneous families, chosen to span the connectivity regimes
that determine partitioning difficulty relative to the separator budget
$k-1=5$: Barab\'asi--Albert graphs~\cite{BA_graph} with attachment degree $d=3$ (vertex
connectivity $\kappa\!\approx\!3 \le k-1$), BA graphs with $d=7$
($\kappa\!\approx\!7 > k-1$, freezing required), dense Erd\H{o}s--R\'enyi
graphs~\cite{random_ER_graph} with mean degree $\langle d\rangle = 12$ (mixed regime), and a
two-cluster BA topology joined by two bridge edges ($\kappa=2$, an easy
control); the random $d$-regular family~\cite{d_regular_graph} is analysed separately in
Sec.~\ref{subsec:regular-scalability}. Instance sizes cover four scale
categories, small
($n=10$--$100$, five seeds), medium ($n=150$--$500$, three seeds), large
($n=600$--$1000$, three seeds), and extreme ($n=2000$--$10{,}000$, two
seeds) for a total of 256 instances and 2{,}616 full decompositions
(every (instance, method) pair across the budget sweep below, including the
DC-QAOA baseline).
The freeze budget is swept over $B_f \in \{1,\dots,10\}$ for $n \le 1000$
and over the representative subset $\{1,2,3,5,10\}$ at the extreme scale;
all budget-resolved aggregates below therefore use the $n\le1000$ subset.
No run exceeded the 90-second wall-clock cap.

\emph{Which partitioner is measured.}
Because the exhaustive separator search of Algorithm~\ref{alg:frozenlgp} is
$\mathcal{O}(n^{k-1})$ and hence intractable at these scales, the
partition-level results in this section
(Tables~\ref{tab:graph-families}--\ref{tab:budget-sweep}, $n$ up to $10^4$) are
produced by a polynomial-time \emph{upper-bound} driver
(\ref{app:driver-agreement}). It constructs feasible cuts but does
not certify that their cardinality is minimal. Consequently, the large-scale
coverage curves and apparent budget thresholds are empirical properties of
that driver on the sampled graph families, not exact statements about all
members of those families. \ref{app:driver-agreement} separately
compares it with the exact production procedure on a moderate-scale audit set
and reports both agreements and disagreements. The conditional
minimum-vertex-cut statement in Sec.~\ref{sec:phase2} applies only to the
exact driver.

We record five groups of preprocessing metrics per run: \emph{coverage}
(fraction of instances successfully decomposed), \emph{runtime}, the node redundancy
level $\mathrm{NRL}=\sum_i |V_i|/n$, \emph{graph reduction} (total frozen
nodes and the reduction ratio $\rho = |F_{\mathrm{total}}|/n$), and the
number of CPP invocations. Furthermore, the details of peak
traced memory, \emph{partition quality} (number of leaf sub-problems,
recursion depth) are reported in~\ref{app:extend_partition_result}.

\emph{Coverage across scales and topologies.}
The baseline behaves exactly as the topological analysis of
~\ref{sec:dcqaoa} predicts: it decomposes the
$\kappa \le k-1$ families (BA $d{=}3$, two-cluster) at every scale, but
fails on essentially all instances whose vertex connectivity exceeds the
separator budget, achieving only 4.6\% coverage on the pooled BA $d{=}7$
and dense-ER instances ($n\le1000$) and 50--54\% coverage overall at every scale category. At $B_f=3$, adding Phase~2 raises coverage without
CPP to 89.8\% on that same pooled high-connectivity subset; the remaining 10.2\% require CPP. Separately, the complete pipeline attains 100\% CPP-inclusive operational coverage in all 2{,}360 FrozenLGP runs at every scale and for every budget $B_f\ge1$ (the 2{,}616 decompositions less the 256 DC-QAOA baseline runs). This full-pipeline result reflects the CPP termination safeguard and is not attributed to freezing alone. The 89.8\% figure is restricted to the $n\le1000$, $B_f=3$ subset; it is not a phase-resolved estimate for all scales. CPP does not promise connected children or exact recovery of omitted cross edges.

\begin{figure*}[t]
    \centering
    \includegraphics[width=0.8\textwidth]{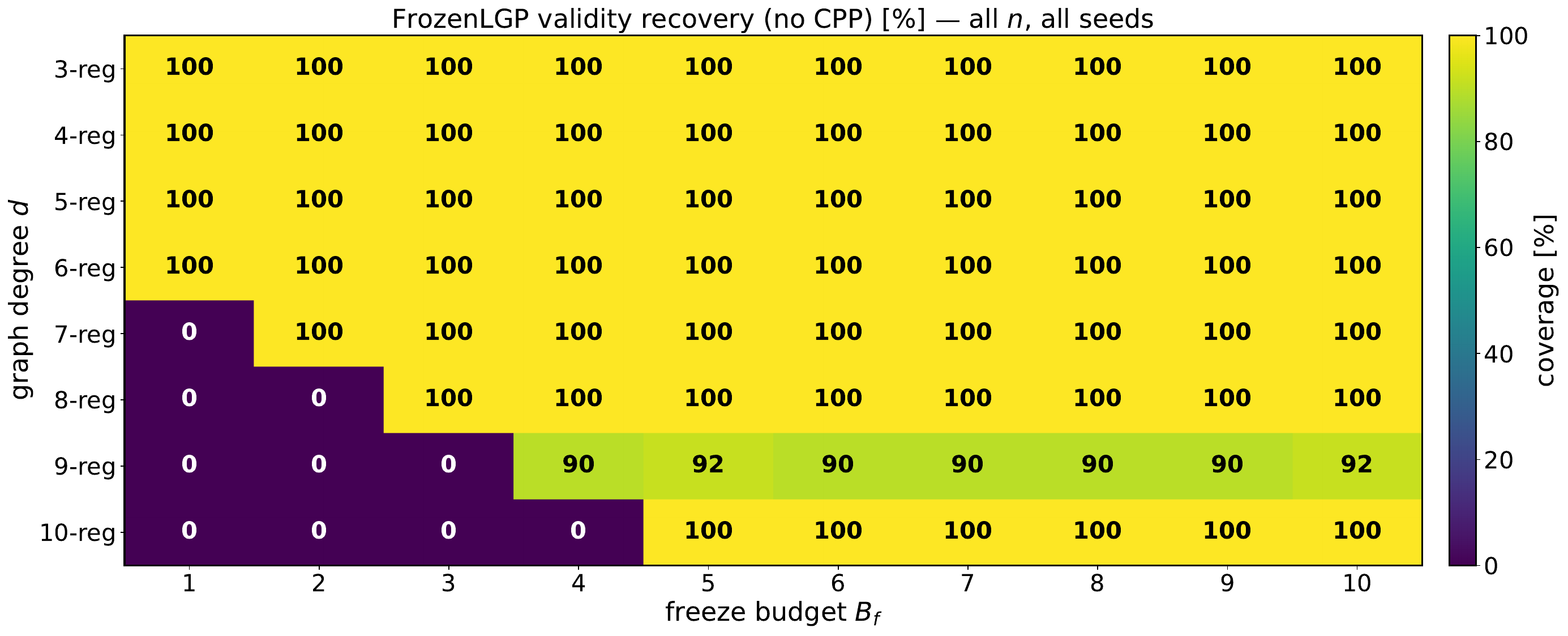}
    \caption{Fraction of random $d$-regular instances decomposed by
    freezing alone (CPP excluded), as a function of degree $d$ and freeze
    budget $B_f$, pooled over all sizes and seeds. Validity recovery
    follows the predicted staircase $B_f \ge d-(k-1)$ on the sampled
    instances, with a binary $0\%\!\to\!100\%$ transition at every degree except $d{=}9$.
    This is an empirical result for the scalable upper-bound driver, not an
    exact theorem for all $d$-regular graphs. The
    off-pattern row at $d{=}9$ ($90-92\%$) is the $n{=}10$
    instance group, for which the 9-regular graph on ten vertices is the
    complete graph $K_{10}$, correctly unpartitionable by any
    vertex-removal strategy and absorbed by the CPP fallback. Including
    CPP, FrozenLGP coverage is 100\% for every $(d, B_f)$ cell.}
    \label{fig:reg-heatmap}
\end{figure*}

\emph{Effect of the freeze budget.}
Fig.~\ref{fig:scal-coverage}(a) and Table~\ref{tab:budget-sweep}
resolve the contribution of the freezing phase itself by counting only
instances recovered \emph{without} CPP. Validity recovery exhibits a
sharp threshold at the topologically expected budget
$B_f = \kappa - (k-1)$: on the hard families, a budget of one recovers
only 9.3\% of instances, a budget of two recovers 79.6\%, and a budget of
three recovers 89.8\%, after which the curve is flat (90.7\% for
$B_f \ge 4$; the residual ${\sim}9\%$ involve at least one sub-problem
whose local connectivity exceeds $k-1+B_f$ at some recursion level and is
absorbed by CPP). The budget also has a pronounced runtime effect below
the threshold: at $B_f{=}1$ the hard-family decomposition averages
0.23\,s because most instances fall through to the (max-flow-probing)
failure path before CPP engages, versus 8-11\,ms for $B_f\ge2$. Beyond the
observed knee, runtime, frozen-node count, reduction ratio, and NRL change
little with $B_f$ in this preprocessing experiment because the driver usually
accepts a smaller feasible cut before exhausting the allowed budget. This does
not make a larger budget free end to end: each accepted set introduces a local
assignment factor, and those factors can multiply across recursion
[Eq.~\eqref{eq:recursive-work}]. The empirical knee at $B_f\in\{2,3\}$
therefore motivates the $m_{\max}{=}3$ default, while aggregate quantum work is
measured separately rather than inferred from the preprocessing curves.

Across 256 instances spanning three orders of magnitude in size and four
connectivity regimes, the scalable driver achieved 100\% CPP-inclusive
operational coverage. Its separator-based recovery is separately measured by the 89.8\% coverage without CPP at $B_f=3$ on the pooled high-connectivity subset ($n\le1000$), versus 4.6\% for Phase~1 alone. Measured preprocessing time was seconds or less and the freezing ratio reached 38-39\%(\ref{app:extend_partition_result}). These statements concern the classical driver and do not imply that the full recursive quantum workload is negligible. We next examine the particularly
regular staircase observed on the random-regular sample.

\subsection{Generalisation to Random Regular Graphs}
\label{subsec:regular-scalability}

Random $d$-regular graphs are the canonical benchmark family for MaxCut
and QAOA studies~\cite{fixed_angle_regular_max_cut, qaoa_origin, qaoa_bound_numerical_max_cut}, and they constitute a uniquely
clean probe of the freezing mechanism: every vertex has degree exactly
$d$, random regular graphs are $d$-connected with high probability, and in contrast to the heavy-tailed and Poisson families of
Sec.~\ref{subsec:partition-scalability}, they contain \emph{no
low-degree fringe} that a vertex-separator search could exploit. When a sampled
graph has vertex connectivity $\kappa=d$, Phase~1 cannot succeed for
$d>k-1$, and a natural budget scale is $d-(k-1)$. This connectivity equality
holds with high probability rather than for every finite random-regular graph,
and the scalable driver does not certify a global optimum. We therefore test,
rather than assume, the resulting staircase over the full degree range
$d \in \{3,\dots,10\}$, i.e.\ from comfortably below to twice the
separator budget, under the identical protocol, driver, scale categories,
and budget sweep as before (the size $n{=}75$ is omitted for parity
reasons, $n$ is incremented by one where $nd$ is odd, and the infeasible
combination $d{=}10$, $n{=}10$ is excluded), for a total of 467 instances
and 4{,}737 decomposition runs with zero timeouts.

An empirical budget staircase.
Fig.~\ref{fig:reg-heatmap} shows the central result as a degree~$\times$
budget heatmap of freezing-only coverage. The baseline boundary is sharp:
the Phase-1-only partitioner decomposes 100\% of instances for
$d \le 5 = k-1$ and essentially none beyond it (1.7\% at $d{=}6$, 0\% for
$d \ge 7$). FrozenLGP converts this hard boundary into a budget staircase
that matches the prediction $B_f^{*} = d-(k-1)$ on this sample at every
degree: $d{=}6$ instances are fully recovered at $B_f{=}1$, $d{=}7$ at
$B_f{=}2$, through $d{=}10$ at $B_f{=}5$, in each case as a binary
$0\%\!\to\!100\%$ transition with no partial-coverage region. Comparing
with Fig.~\ref{fig:scal-coverage}(a) clarifies the behaviour observed on
the heterogeneous families: there, the degree \emph{distribution} smooths
the same underlying threshold into a knee (80--90\% recovery around
$B_f \approx \kappa-(k-1)$), whereas the uniform degree of regular graphs
exposes it as a step function. For the sampled random-regular instances,
$B_f=d-(k-1)$ is therefore a useful empirical starting rule. It is neither
necessary nor sufficient in general: finite instances can have $\kappa<d$,
the selected separator matters, and a recursive child can have different
connectivity.

Taken together, the regular-graph study shows that the freezing budget is an
interpretable control knob for the sampled families. The observed coverage,
reduction, and runtime trends (\ref{subsec:reduce_redun_runtime_regular})
extend empirically from heterogeneous scale-free and random topologies to the
uniform-degree instances commonly used in MaxCut benchmarking.
%The quality of the quantum optimisation built on these decompositions is evaluated next.

\subsection{End-to-End Optimization Quality}
\label{subsec:e2e-quality}

A decomposition layer is only worth having if it is invisible where the
pipeline already works and useful where it does not. The end-to-end
evaluation therefore addresses two complementary questions: does FrozenLGP
degrade quality on DC-QAOA-solvable instances, and does it deliver
acceptable quality on previously unsupported instances?

\begin{table}[H]
\caption{\label{tab:e2e-main}%
End-to-end MaxCut results on 100 instances ($n \in \{8,\ldots,20\}$,
noiseless; failures counted as AR\,$=0$; brackets: bootstrap 95\,\% CI).
The two upper blocks are classical and no-decomposition quantum reference
solvers, all of which are applicable at these small sizes and against which the
divide-and-conquer block must be read: at $n\le 20$ the classical heuristics
are strong, so the comparison probes coverage and quality preservation, not
quantum advantage. AR$^{\dagger}$ is each method's AR on \emph{its own} solved
set (set size in parentheses); for methods with full coverage it equals the AR
column. The final column is the like-for-like comparison on the 76 instances
solved by \emph{both} DC-QAOA and FrozenLGP (``---'' where not applicable), on
which the strict phase-priority guarantee makes FrozenLGP and DC-QAOA run
identical circuits. FrozenLGP coverage is CPP-inclusive; without CPP it is 91\% ($B_f = 3$) and 89\% ($B_f = 2$).}
\centering
\resizebox{\textwidth}{!}{%
\begin{tabular}{lllll}
\hline
Method & Coverage & AR [95\,\% CI] & AR$^{\dagger}$ (own set) & AR (76 common)\\
\hline
\multicolumn{5}{l}{\emph{Classical references ($n\le 20$)}}\\
Goemans--Williamson SDP~\cite{goemans_williamson}   & 100\,\%          & 1.000 [1.000, 1.000]          & 1.000 (100) & --- \\
Random search                                & 100\,\%          & 0.994 [0.992, 0.997]          & 0.994 (100) & --- \\
Local search                                 & 100\,\%          & 0.994 [0.990, 0.997]          & 0.994 (100) & --- \\
\hline
\multicolumn{5}{l}{\emph{Quantum, no decomposition}}\\
Monolithic QAOA                              & 100\,\%          & 0.982 [0.974, 0.990]          & 0.982 (100) & --- \\
FrozenQubits~\cite{frozenQubits} ($m{=}3$)   & 100\,\%          & 0.986 [0.981, 0.991]          & 0.986 (100) & --- \\
\hline
\multicolumn{5}{l}{\emph{Divide-and-conquer}}\\
DC-QAOA (Phase 1) & 76\,\% & 0.750 [0.662, 0.828] & 0.986 (76) & 0.986 \\
FrozenLGP ($B_f{=}2$) & \textbf{100\,\%} & 0.985 [0.979, 0.990] & 0.985 (100) & 0.986 \\
FrozenLGP ($B_f{=}3$) & \textbf{100\,\%} & 0.983 [0.977, 0.988] & 0.983 (100) & 0.986 \\
\hline
\end{tabular}
}
\end{table}

\emph{Quality preserved on DC-QAOA-solvable instances.}
We report non-parametric bootstrap 95\% confidence intervals over problem
instances~\cite{bootstrap_test}. The central comparison is paired by graph and
restart: graph generation, optimiser seeds, QAOA parameter initialisations,
shots, node order, and circuit fingerprints are identical across methods. On a
graph for which Phase~1 succeeds, strict phase priority prevents FrozenLGP from
entering Phases~2 or 3. The audit confirms this implementation invariant on all
76 common instances: the selected restart, reconstructed candidate set, best
cut, and approximation ratio are identical pairwise. Equality here is therefore
an exact code-path property of the paired run, not an inference from overlapping
confidence intervals or a non-significant hypothesis test.

\emph{Extended applicability on previously unsupported instances.}
We evaluate the complete pipeline on 100 fixed benchmark instances
with $n \in \{8,10,12,16,20\}$ (BA $d{\in}\{2,3\}$, connected ER
$p{\in}\{0.5,0.8\}$, five seeds each), using exact exhaustive ground truth.
The $n\le20$ range is chosen because it is precisely the regime in which
\emph{every} compared method is simultaneously executable, monolithic
(non-decomposed) QAOA and the FrozenQubits~\cite{frozenQubits} hub-freezing baseline still fit the
register, the classical references (Goemans--Williamson SDP, random and local
search) apply, and exact exhaustive ground truth is attainable, so the dense-graph
quality of all methods can be read off the same instances against the true
optimum. It is a controlled comparison range, not a ceiling on FrozenLGP's
applicability; the larger, decomposition-only regime is studied separately below
and in Sec.~\ref{subsec:qaoa2}.

We compare standard DC-QAOA against FrozenLGP
with $B_f \in \{2,3\}$ and the CPP fallback enabled, and additionally report
four reference solvers that are applicable at these small sizes: the
Goemans--Williamson SDP relaxation~\cite{goemans_williamson}, with the SDP
solved by SCS and 100 seeded hyperplane rounds; an unstructured search over
10{,}000 seeded random assignments; a one-exchange local search with 10 seeded
restarts; and
monolithic (non-decomposed) QAOA together with the FrozenQubits hub-freezing
baseline~\cite{frozenQubits}.
Both methods use $k{=}6$, $p{=}1$, $t{=}20$, 10{,}000 shots,
COBYLA~\cite{cobyla} with at most 300 iterations, and three independent restarts per
instance (best cut reported); noiseless simulations use analytic Qiskit~\cite{qiskit_tool}
Aer statevector expectation values, whereas the noise study uses shot-based
estimation under the active noise model
(software versions and transpilation settings in \ref{sec:noise_rubus_validation}).
Approximation ratios count failures as zero and bootstrap intervals resample
problem instances. A separate noise study, with two declared
channels and one deterministic restart, is reported in
\ref{sec:noise_rubus_validation}; its distributional scope differs and
is not pooled with this noiseless table.

Of the 100 instances, 76 are ones on which DC-QAOA executes
normally.
On these 76 instances, FrozenLGP applies zero freezing because Phase~2 is
never entered when Phase~1 succeeds. Table~\ref{tab:e2e-main} therefore reports
the same common-set mean, AR $0.9863066$, for DC-QAOA and for both freeze
budgets. All 456 paired provenance/invariance checks pass (76 instances,
two budgets, and three audited quantities per pair). Statistical equivalence
testing is unnecessary for an exactly zero paired difference, the result is scoped to the shared Phase-1-success set.

The remaining 24 instances, predominantly dense ER ($p{=}0.8$) and
BA ($d{=}3$) graphs at $n{=}20$, lie outside the
regime: DC-QAOA returns failure and contributes AR\,$=0$ to its
failure-inclusive mean, pulling it to $0.750$ overall.
FrozenLGP recovers these instances through MVC freezing (Phase~2; 15 best-restart outcomes at $B_f = 3$) and CPP fallback (Phase 3; 9 outcomes) then achieves full-pipeline failure-inclusive AR $0.983\,[0.977,0.988]$. At $B_f =2$, the corresponding phase counts are 76, 13, and 11 and AR is
$0.985\,[0.979, 0.990]$. These results demonstrate empirical coverage extension.

\emph{Classical and single-circuit references.}
At $n\le 20$ every instance is small enough that classical and non-decomposed
solvers apply, and Table~\ref{tab:e2e-main} reports them so that the quantum
decomposition results can be read in context. The Goemans--Williamson
SDP~\cite{goemans_williamson} attains the optimum on all 100 instances
(AR $1.000$), and even unstructured random search and one-exchange local search
reach AR $\approx0.994$; monolithic QAOA and FrozenQubits attain AR $0.982$ and
$0.986$. FrozenLGP is therefore \emph{not} competitive with strong classical
heuristics at these sizes, nor is it intended to be: the small-$n$ regime is
precisely where classical MaxCut is easy. The purpose of this benchmark is to
establish, against exact ground truth, that converting a structurally infeasible
instance into valid sub-problems costs no solution quality relative to DC-QAOA,
and that the decomposition stays within the quality tier of single-circuit QAOA.
The regime in which decomposition is \emph{necessary}, graphs wider than the
available register, where monolithic QAOA, FrozenQubits, and (on dense inputs)
DC-QAOA cannot be executed at all, is the subject of the partition-level
scalability study of Sec.~\ref{subsec:partition-scalability} and the resource
analysis of~\ref{subsec:e2e-resources}. The $n\le20$ cap in this table
is therefore a property of the \emph{reference} solvers, not of the partition
driver. Classical preprocessing is evaluated to $n=10^4$, whereas optimization
quality uses exact ground truth only where the reference solver certifies an
optimum. The at-scale QAOA-in-QAOA comparison in Sec.~\ref{subsec:qaoa2}
retains results through $n=100$ and labels the dense $n=100$ denominator as a
best-known reference. Its event-level total-resource audit stops at $n=20$; no
exact gate--shot total is extrapolated from the large-$n$ invocation proxy.
\ref{sec:noise_rubus_validation} reports the separate simulated-noise
stress test and explicitly distinguishes best-sample from distributional metrics.

\begin{table*}[t]
\caption{\label{tab:qaoa2}%
At-scale FrozenLGP--QAOA\textsuperscript{2} comparison under aligned
per-subproblem settings ($k=6$, $p=1$, $10^4$ sampling shots, COBYLA maximum
300 evaluations). AR and $N_{\mathrm{inv}}$ are means; $N_{\mathrm{inv}}$ is
the observed number of invoked subcircuit solves and hence a first-order total
work proxy, not an exact shot or gate--shot count. Both methods return a
solution on every listed instance. Sparse rows pool three seeds from each of BA
$d=3$ and random 4-regular graphs. Random-regular rows use five seeds. References
are solver-certified CP-SAT~\cite{cpsatlp} optima except for the dense $n=100$ block, which uses
the better of Goemans--Williamson~\cite{goemans_williamson} and local-search values.}
\centering
\begin{tabular}{lrrrr}
\hline
Scope & FrozenLGP AR & QAOA\textsuperscript{2} AR & FrozenLGP $N_{\mathrm{inv}}$ & QAOA\textsuperscript{2} $N_{\mathrm{inv}}$\\
\hline
\multicolumn{5}{l}{\emph{Sparse scaling: BA $d=3$ and random 4-regular (six instances per size)}}\\
$n=24$ & \textbf{0.963} & 0.850 & 18.2 & 5.0 \\
$n=32$ & \textbf{0.974} & 0.856 & 26.0 & 7.0 \\
$n=50$ & \textbf{0.921} & 0.821 & 42.5 & 12.0 \\
$n=75$ & \textbf{0.926} & 0.763 & 66.7 & 17.0 \\
$n=100$ & \textbf{0.937} & 0.759 & 90.3 & 21.0 \\
\hline
\multicolumn{5}{l}{\emph{Random $d$-regular, $n=50$ (exact; five instances per degree)}}\\
$d=5$ ($B_f=0$) & \textbf{0.941} & 0.805 & 43.0 & 12.0 \\
$d=6$ ($B_f=1$) & \textbf{0.939} & 0.802 & 42.6 & 12.0 \\
$d=7$ ($B_f=2$) & \textbf{0.918} & 0.795 & 116.0 & 12.0 \\
$d=8$ ($B_f=3$) & \textbf{0.949} & 0.808 & 812.0 & 12.0 \\
\hline
\multicolumn{5}{l}{\emph{Random $d$-regular, $n=100$ (best-known reference; five instances per degree)}}\\
$d=5$ ($B_f=0$) & \textbf{0.971} & 0.766 & 90.6 & 21.0 \\
$d=6$ ($B_f=1$) & \textbf{0.956} & 0.772 & 91.4 & 21.0 \\
$d=7$ ($B_f=2$) & \textbf{0.956} & 0.812 & 319.6 & 21.0 \\
\hline
\end{tabular}
\end{table*}

\begin{figure*}[t]
    \centering
    \includegraphics[width=0.8\textwidth]{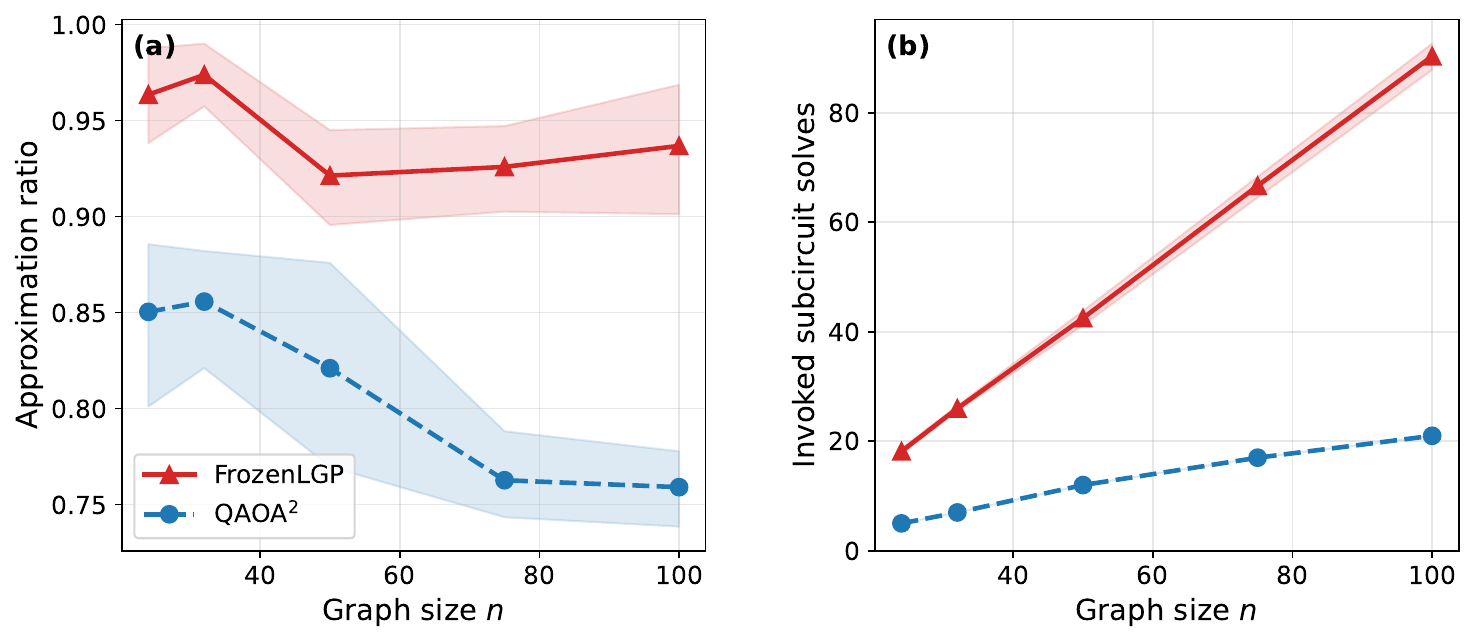}
    \caption{Sparse at-scale comparison under aligned per-subproblem settings.
    (a) Approximation ratio and (b) invoked subcircuit solves, the recorded
    first-order work proxy. Markers and bands are pooled means and bootstrap
    95\% confidence intervals over six instances at each size (three seeds from
    each of two graph families). Equal per-subproblem settings make solve count
    informative about scaling, but the counter-level audit in
    Table~\ref{tab:qaoa2-resources} is required for exact total-shot and
    two-qubit-gate--shot exposure.}
    \label{fig:qaoa2}
\end{figure*}

\subsection{At-Scale Comparison with QAOA-in-QAOA}
\label{subsec:qaoa2}

QAOA-in-QAOA (QAOA\textsuperscript{2})~\cite{qaoa-in-qaoa} provides a useful
full-coverage decomposition reference because both methods recursively reduce a
large graph to width-bounded QAOA subproblems. The comparison therefore retains
the original scale range through $n=100$. Both implementations use the same
per-subproblem settings: $k=6$, $p=1$, $10^4$ sampling shots, and COBYLA with a
maximum of 300 objective evaluations. These settings align each invoked solve,
but do not align the number of solves or total quantum work.

Table~\ref{tab:qaoa2} and Fig.~\ref{fig:qaoa2} show the quality--work trade-off
that was hidden by reporting approximation ratio alone. On the sparse arm,
FrozenLGP has higher mean AR at all five measured sizes. At $n=100$, the means
are 0.937 versus 0.759, while FrozenLGP invokes 90.3 subcircuit solves compared
with 21.0 for QAOA\textsuperscript{2}. Thus the observed quality difference at
that size is accompanied by approximately $4.3\times$ as many invoked solves;
it is not an equal-total-cost result. The dense degree sweep has the same AR
ordering on the sampled instances, but also makes the cost sensitivity visible:
at $n=50,d=8$, FrozenLGP invokes mean 812 solves compared with 12 for
QAOA\textsuperscript{2}. The measured ordering is reported as observed; it
does not by itself establish that the AR gap widens with scale or that either
method is universally preferable.

\begin{table*}[t]
\caption{\label{tab:qaoa2-resources}%
Counter-level resource audit on 18 exact-ground-truth instances (BA $d=3$ and
random 7-regular; $n\in\{12,16,20\}$; three seeds). ``Default'' uses the
implementations' original stopping settings. ``Matched'' imposes the same
per-instance caps of 192 objective evaluations and 65{,}536 total shots. Values
are means. ``Opt.'' is the number of variational optimisations; ``unique'' is
the number of distinct parameterised circuit structures; two-qubit-gate--shots
sum two-qubit gates times shots over all circuit executions; ``classical'' is
decomposition, coarsening/merging, and reconstruction time (all recorded
non-quantum phases). Both methods return a solution on every instance.}
\centering
\resizebox{\textwidth}{!}{%
\begin{tabular}{llrrrrrrr}
\hline
Setting & Method & AR & Opt. & Obj. evals & Total shots & Unique & 2q gate--shots ($10^6$) & Classical (s)\\
\hline
Default & FrozenLGP & 0.943 & 20.8 & 627.6 & 101,611 & 18.6 & 1.206 & 2.169 \\
Default & QAOA\textsuperscript{2} & 0.924 & 4.0 & 107.9 & 17,906 & 4.0 & 0.164 & 0.001 \\
Matched & FrozenLGP & 0.952 & 20.8 & 186.3 & 65,431 & 18.6 & 0.759 & 2.208 \\
Matched & QAOA\textsuperscript{2} & 0.915 & 4.0 & 105.2 & 54,506 & 4.0 & 0.497 & 0.001 \\
\hline
\end{tabular}
}
\end{table*}

In these implementations each invoked subcircuit solve launches one variational
optimisation, so $N_{\mathrm{inv}}$ directly tracks the first requested resource
counter. It cannot reconstruct every other counter exactly: optimisers can stop
early, estimator-shot allocations differ by experimental arm, and circuit edge
counts vary. Table~\ref{tab:qaoa2-resources} therefore reports the full counters
where event-level telemetry is available. Under the original settings,
FrozenLGP uses mean 20.8 optimisations, 627.6 objective evaluations, 101{,}611
shots, 18.6 unique structures, 1.206 million two-qubit-gate--shots, and 2.169 s
of recorded non-quantum phase time, compared with 4.0, 107.9, 17{,}906,
4.0, 0.164 million, and 0.001 s for QAOA\textsuperscript{2}. Under the common
caps, mean AR is 0.952 versus 0.915, but the realised resources are still
unequal: 0.759 versus 0.497 million gate--shots. The small counter audit supports
using invocation count as a scale indicator while also showing why it must not
be relabelled as an exact total quantum cost.

\section{Discussion}
\label{sec:discussion}

The results support FrozenLGP as a recovery layer for the
separator-dependent LGP/MDR path of DC-QAOA, rather than as a replacement for every decomposition method. Strict phase priority leaves the original path unchanged whenever LGP succeeds, as verified by the 76 paired common-path instances. At $B_f=3$, adaptive freezing raises separator-based coverage without CPP from 4.6\% to 89.8\% on the pooled high-connectivity partition benchmarks ($n\le1000$). In the exact-ground-truth end-to-end arm, it raises coverage without CPP from 76 to 91 of 100 instances: 76 selected best-restart
outcomes use Phase~1 and 15 additional outcomes use Phase~2. CPP handles the remaining nine outcomes, giving 100\% CPP-inclusive coverage and full-pipeline mean AR 0.983. Likewise, the 100\% operational coverage observed in the partitioning study up to $10^4$ vertices includes CPP. The freezing-based recovery is the central method result; CPP supplies a distinct termination
guarantee. Operational coverage means that the configured pipeline returns a candidate, without guaranteeing connected fragments, preserved edges, approximation quality, or efficient total execution.

\subsection{Quality Preservation versus Invocation Cost}
\label{subsec:discussion-tradeoff}

The comparison with QAOA-in-QAOA~\cite{qaoa-in-qaoa} exposes two different
responses to the same full-coverage requirement. QAOA-in-QAOA permits
arbitrary width-bounded blocks
and pays the price of a lossy cut-edge merge: after local block solutions are
chosen, crossing-edge information is compressed into classically computed
coarse couplings and is resolved through merge-level variables rather than
being optimized jointly in the original quantum subcircuits. FrozenLGP takes
the opposite route in Phases~1 and~2. It preserves a shared separator and, for
each selected frozen assignment, folds the incident boundary couplings exactly
into local bias fields on the active qubits. In this conditional sense the
boundary terms are retained rather than discarded, while only the obstructing
vertices in the bounded freeze budget are absorbed as local frozen-spin biases.
This exact statement concerns
the conditional Hamiltonian substitution; assignment choice, variational
optimization, top-$20$ MDR truncation, and final sample selection remain
approximate. Phase~3 is also an explicit exception because its arbitrary split omits cross edges from the child Hamiltonians.

Both methods achieve 100\% operational coverage in the at-scale comparison;
for FrozenLGP this is CPP-inclusive coverage, distinct from the 89.8\% partition-level coverage without CPP ($n \leq 1000, B_f = 3$) and the 91/100 end-to-end coverage without CPP summarized above for their respective benchmark sets. The at-scale experiment compares the full pipelines' sampled solution quality under the tested per-subproblem settings.
FrozenLGP has higher
mean AR in every reported sparse-size and dense-degree row, but it invokes more
subcircuit solves. At sparse $n=100$, the AR values are 0.937 and 0.759 while
the mean invocation counts are 90.3 and 21.0 for FrozenLGP and
QAOA-in-QAOA, respectively. The contrast becomes larger in the dense
$n=50,d=8$ case, where FrozenLGP invokes 812 solves and QAOA-in-QAOA invokes
12. QAOA-in-QAOA therefore occupies the lower-invocation, lower-AR side of the
observed trade-off, whereas FrozenLGP occupies the higher-invocation, higher-AR
side. Neither method dominates both axes. The appropriate choice depends on
whether the application values the additional approximation quality enough to
justify the larger quantum workload.

The counter-level arm reinforces this interpretation without making the two
methods exactly equal in realized cost. Under shared caps, FrozenLGP attains AR
0.952 versus 0.915, but uses 0.759 million two-qubit-gate--shots versus 0.497
million. Invoked solves are consequently useful as a first-order scaling proxy
when $k$, $p$, shots per solve, and the optimizer limit are aligned, but they
are not an exact quantum-cost measure. Optimizers can terminate at different
evaluation counts, circuit structures contain different numbers of two-qubit
gates, and estimator and sampling shots contribute differently to the total.
The comparison establishes a finite quality--work frontier, not a universal
advantage for either decomposition strategy.

\subsection{Limitations and Scope}
\label{subsec:discussion-limitations}

The first group of limitations concerns the partition and reconstruction
algorithms. Phase~2 returns a minimum node cut only conditional on the current
separator candidate and accepts the first feasible pair; it does not jointly
minimize the frozen set over all separators. In the driver-agreement audit of
\ref{app:driver-agreement}, the production
exact driver matches the global-over-separators reference in 51 of 69 comparable
cases, and minimizing cardinality does not by itself minimize the induced field
strength or the quality effect of fixing those spins. The scalable and figure
drivers are polynomial-time empirical approximations rather than
global-optimality certificates. If both separator-compatible phases fail, CPP
guarantees only a
width-feasible continuation. Its children may be disconnected, and on the
CPP-triggered stress cases the omitted cross edges contain mean 52.4\% and
maximum 69.2\% of the parent edge weight. Full-parent-objective ranking improves
the choice among retained candidates but cannot restore probability mass or
bitstrings removed before the top-$20$ ranking stage. Accordingly, neither MDR
nor CPP reconstruction is exact end to end.

The second group concerns computational resources. The six-qubit leaf-width
bound and the per-circuit depth and gate reductions do not bound aggregate
work. Conditional assignments multiply along the recursion tree, reaching
$1.17\times10^6$ execution inflation in the recursive-work audit of
Table~\ref{tab:recursive-work}. Split imbalance is a separate central
limitation: the partition audit finds mean balance $\beta<0.01$ at $n=10^4$,
so minimum-cut peeling produces an approximately linear-depth sequential chain
rather than a logarithmic-depth parallel tree. The reported circuit-resource comparison assumes all-to-all
connectivity and excludes device routing, queueing, and calibration overhead.
Classical preprocessing timings are CPU- and implementation-dependent.
No quantitative comparison with wire-cutting methods such as
CutQC~\cite{cutQC} is reported, because the two cost models are structurally
incommensurable: the CutQC classical-postprocessing factor $4^K$ is governed
by the number of cut wires $K$ of one selected cut on one circuit, whereas
FrozenLGP's total work follows the multiplicative recurrence of
Eq.~\eqref{eq:recursive-work}, in which assignment factors compound along the
recursion tree and depend on the frozen-set sizes realized at every internal
node. A single local factor from either model says nothing about the other's
end-to-end cost, which additionally depends on cut selection, observable
grouping, backend, optimizer budget, and reconstruction implementation.
Any such comparison would require matched end-to-end execution on common
instances, which is left outside the present scope. These results therefore
establish bounded
circuit width, not bounded total quantum work or hardware wall-clock advantage.

Finally, the empirical objective scope is deliberately narrow. The folding
identity extends to weighted edge-only MaxCut and to pairwise Ising couplings,
but weighted instances and broader Ising/QUBO objectives are not empirically
validated here. Initial or inherited one-body fields also remove the global
spin-flip symmetry used to halve some frozen-assignment evaluations.

\subsection{Future Directions}
\label{subsec:discussion-future}

Future work should treat solution quality and aggregate resource use as a
joint optimization problem. Promising directions include separator selection
that balances child sizes and boundary information, pruning or sharing
conditional optimizations across recursion, and reconstruction rules that
retain more cross-partition information before candidate truncation.
One structural reduction of the conditional workload appears particularly
promising. The $2^{|F|}$ assignment factor could be reduced by
landscape-similarity-guided optimization~\cite{do_qaoa}: the conditional
sub-circuits of a partition differ only in their local bias fields, so their
cost landscapes are closely related, and optimized parameters can be
transferred or shared across assignments instead of re-optimizing each of the
$2^{|F|}$ variants independently.
Physical-device experiments with hardware-derived noise parameters, routing,
queueing, and wall-clock accounting are required to determine where the
additional FrozenLGP subcircuit work is justified by its observed quality gain.

\section{Conclusion}
\label{sec:conclusion}

FrozenLGP provides an adaptive recovery policy for the separator-dependent
DC-QAOA pipeline. It retains standard LGP when possible, uses conditional
qubit freezing to recover additional separator-compatible cases, and supplies
a deterministic width-feasible fallback when both separator paths fail. The
fixed-spin folding identity preserves incident couplings in each conditional
Hamiltonian, allowing obstructing vertices to be removed from the active
quantum register while their assignment-dependent influence remains encoded.

At $B_f=3$, adaptive freezing raises coverage without CPP from 4.6\% to
89.8\% on the pooled high-connectivity partition benchmarks ($n\le1000$).
In the 100-instance end-to-end benchmark, it extends the separator-based
route from 76 Phase-1 outcomes to 91 outcomes including 15 recovered by
Phase~2, using the selected best-restart classification. CPP handles the
remaining nine outcomes. The resulting 100\% CPP-inclusive operational
coverage, also observed in the partitioning study up to $10^4$ vertices, is
a termination guarantee rather than a freezing-based coverage contribution.
The common Phase-1 execution path remains unchanged.

Relative to QAOA-in-QAOA,
FrozenLGP yields higher approximation ratios on the reported at-scale cases at
the price of more subcircuit invocations and greater aggregate quantum work.
The principal result is therefore a quality--work design option for
width-limited divide-and-conquer QAOA: separator preservation can improve
solution quality when the available execution budget can support the added
conditional subcircuits.
This width guarantee does not imply a shallow decomposition: at $n=10^4$ the
observed mean split balance falls below 0.01, producing approximately linear
rather than logarithmic recursion depth and a long sequential execution path.

\ack{This work was partly supported by Institute for Information \& communications Technology Planning \& Evaluation (IITP) grant funded by the Korea government (MSIT) (No. 2020-0-00014, A Technology Development of Quantum OS for Fault-tolerant Logical Qubit Computing Environment) and Creation of the quantum information science R\&D ecosystem(based on human resources) through the National Research Foundation of Korea(NRF) funded by the Korean government (Ministry of Science and ICT(MSIT)) (No. RS-2023-00256050).}

\funding{Institute for Information \& communications Technology Planning \& Evaluation, Grant ID: 2020-0-00014 and National Research Foundation of Korea, Grant ID: RS-2023-00256050}
% This section is a list of funder names and grant numbers

\roles{S.S. conceived the original idea, designed the methodology, implemented the algorithms, conducted the experiments, analyzed the results, and wrote the manuscript. L.H., D.K. provided guidance on data visualization, contributed feedback on the experimental results, and assisted during manuscript preparation. Y.H. supervised the project, provided scientific guidance, suggested improvements to the methodology and manuscript, and contributed to the overall development of the work. All authors reviewed and approved the final manuscript.}
% List author names and the contributions made to the article, using terms from the NISO Contributor Roles Taxonomy (CRediT) https://credit.niso.org

\data{The source code supporting this study is publicly available via GitHub, with version 1.0.0 released at \url{https://github.com/QCL-PKNU/FrozenLGP/releases/tag/v1.0.0}.}

\appendix
\section{Divide-and-Conquer QAOA and the Large Graph Partitioner}
\label{sec:dcqaoa}

The variational ansatz introduced in Sec.~\ref{sec:prelim_qaoa_mapping} maps every vertex of the input
graph to a physical qubit, so the largest MaxCut instance that can be solved
by a monolithic QAOA circuit is hard-bounded by the qubit register width of
the underlying device. On near-term hardware~\cite{qubit_mapping_nisq, swap_gates, hardware_aware_mapping} this width is small, typically
a few tens of high-quality qubits while the graphs of practical interest in
network analysis, logistics, and clustering routinely contain hundreds of
nodes. Divide-and-Conquer QAOA (DC-QAOA), introduced by
Li~\textit{et~al.}~\cite{dc-qaoa}, addresses this gap by recursively
decomposing a large graph $G=(V,E)$ into sub-graphs that respect a
user-specified qubit budget $k$, solving each leaf sub-graph with an
independent QAOA sub-circuit, and classically reconstructing the global
solution distribution from the per-leaf samples. The framework therefore
trades a single deep circuit on a wide register for a controlled number of
shallow circuits on a narrow register, the regime in which NISQ devices are
known to deliver their highest fidelity. DC-QAOA rests on two algorithmic
primitives, the Large Graph Partitioner (LGP) and the Measurement
Distribution Reconstruction (MDR), and on a single efficiency metric, the
Node Redundancy Level (NRL); we summarise each in turn.

\subsection{Large Graph Partitioner (LGP)}\label{dc_qaoa_LGP}
The LGP policy is responsible for splitting any graph that exceeds the qubit
budget into two sub-graphs that each fit. Given a graph $G=(V,E)$ with
$|V|=n>k$, LGP iterates over candidate \emph{separator sets} $S\subseteq V$
in increasing order of size $|S|=1,2,\dots,k-1$ and, for each $S$, removes
the corresponding vertices to obtain the residual graph $G\setminus S$. The
separator $S$ is declared \emph{valid} if and only if
\begin{equation}
\label{eq:lgp-condition}
\bigl|\,\textsc{ConnectedComponents}(G\setminus S)\,\bigr|=2,
\end{equation}
i.e. the removal cleanly disconnects $G$ into exactly two non-empty
sub-components $V_1'$ and $V_2'$ with no cross-component edges (Fig.~\ref{fig:frozenlgp_overview} (a-b)). When
Eq.~\eqref{eq:lgp-condition} is satisfied, LGP returns the two sub-graphs
\begin{equation}
\label{eq:lgp-subgraphs}
G_1 = G\bigl[V_1'\cup S\bigr],\qquad
G_2 = G\bigl[V_2'\cup S\bigr],
\end{equation}
each augmented by the shared separator vertices $S$ so that the boundary
information between the two parts is preserved. If $|V_1\cup S|>k$ or
$|V_2\cup S|>k$, LGP is invoked recursively on the offending sub-graph; the
recursion terminates when every leaf sub-graph satisfies $|V_i|\leq k$, at
which point the leaves are dispatched to the QAOA solver of Sec.~\ref{sec:prelim_qaoa_mapping}. The
DC-QAOA framework therefore inherits the entire variational machinery, the
parameterised cost-mixer ansatz, the classical optimisation
loop, and the measurement step unchanged at the leaves; the only addition
is the recursive decomposition managed by LGP. Fig.~\ref{fig:frozenlgp_overview}(a,b) illustrates a
successful LGP partition of a sparse graph using nodes $\{4,5\}$ as the
separator.

A critical implicit assumption of Eq.~\eqref{eq:lgp-condition} is that some
separator $S$ of size at most $k-1$ actually exists. This topological
prerequisite is satisfied for sparse graphs with low vertex connectivity but
breaks down systematically on dense graphs, where the residual
$G\setminus S$ remains connected for every $S$ of admissible size. When no
valid separator can be identified, LGP terminates with a partition failure
and the entire DC-QAOA pipeline aborts without
returning any candidate bitstring. This failure mode is the central
motivation for the present work and is quantified empirically in
Section~\ref{sec:evaluation}.

\subsection{Measurement Distribution Reconstruction (MDR)}\label{dc_qaoa_MDR}
Once LGP has produced a pair of leaf sub-graphs, each sub-graph is solved by
an independent QAOA sub-circuit and yields a measurement distribution
$D_i=\{(b_i,c_i)\}$ over $|V_i|$-bit bitstrings, where $c_i$ is the sample
count of bitstring $b_i$. The Measurement Distribution Reconstruction (MDR)
policy stitches $D_1$ and $D_2$ into a parent distribution $D_P$ over the
$n$ vertices of the original graph through three steps.

First, MDR enforces a \emph{separator compatibility} predicate. Because the
separator vertices $S$ appear in both sub-graphs, a pair of bitstrings
$(b_1,b_2)\in D_1\times D_2$ is admissible only if it agrees on the bits
indexed by $S$:
\begin{equation}
\label{eq:mdr-compat}
b_1[i_{S}^{G_1}] \;=\; b_2[i_{S}^{G_2}]
\qquad
\forall\,\text{separator node in }S,
\end{equation}
where $i_{S}^{G_j}$ is the position of a given separator vertex in the
ordered node list of sub-graph $G_j$. Pairs that disagree on any separator
bit are discarded as physically inconsistent. Second, each admissible pair
is assembled into a parent bitstring by concatenating $b_1$ with the
$G_2$-exclusive portion of $b_2$ in a stable left-to-right node order,
producing a single $n$-bit string $b$ over the union $V_{G_1}\cup V_{G_2}$.
Third, every admissible pair contributes a non-negative weight to
$D_P[b]$ under one of three scoring schemes:
% \begin{equation}
% \label{eq:mdr-schemes}
% \phi_{\text{min}}(c_1,c_2)=\min(c_1,c_2),\quad
% \phi_{\text{mul}}(c_1,c_2)=c_1c_2,\quad
% \phi_{\text{minXmul}}(c_1,c_2)=\min(c_1,c_2)\,c_1c_2.
% \end{equation}
\begin{equation}
\label{eq:mdr-schemes}
\begin{aligned}
\phi_{\text{min}}(c_1,c_2) &= \min(c_1,c_2), \\
\phi_{\text{mul}}(c_1,c_2) &= c_1c_2, \\
\phi_{\text{minXmul}}(c_1,c_2) &= \min(c_1,c_2)\,c_1c_2.
\end{aligned}
\end{equation}
The \emph{min} scheme is conservative and favours balanced sub-circuit
support; the multiplicative scheme is aggressive and amplifies the
high-probability modes of both sub-distributions; the hybrid
\emph{minXmul} scheme balances the two by penalising configurations whose
two contributing sub-circuit counts are heavily skewed while still rewarding
joint high probability. After scoring, the parent distribution is sorted,
truncated to the top $t$ states, and rescaled to the requested shot-count
total. This is a heuristic reconstruction of retained candidates, not an
exact recovery of the joint quantum distribution. We use $t=20$ and the
\emph{minXmul} rule throughout the reported DC-QAOA and FrozenLGP comparisons.

\subsection{Node Redundancy Level (NRL)}\label{dc_qaoa_NRL}
LGP duplicates the separator vertices $S$ across both sub-graphs in order to
preserve boundary edge information; this duplication grows the cumulative
vertex count submitted to the QAOA solver beyond the original $n$. The
Node Redundancy Level~\cite{dc-qaoa} quantifies this overhead. For a
single LGP step that returns the pair $(G_1,G_2)$ from a parent graph $G$,
it is defined as
\begin{equation}
\label{eq:nrl}
\text{NRL}\bigl(G_1,G_2 \mid G\bigr)
\;=\;\frac{|V_{G_1}|\,+\,|V_{G_2}|}{|V_G|}
\;=\;1\,+\,\frac{|S|}{|V_G|}.
\end{equation}
For a completed recursive decomposition we report the operational aggregate
$\mathrm{NRL}=\sum_{\ell\in\mathcal L}|V_\ell|/|V_G|$, where $\mathcal L$
is the set of leaf circuit structures. It measures duplicated leaf width,
not repeated executions caused by frozen assignments. Lower NRL can
therefore coexist with a larger total shot or optimizer budget; those
resources are reported separately.

Crucially, NRL captures only the cost incurred when LGP is able to find
\emph{some} valid separator. It is silent about partition failures: when
LGP fails to partition the graph, NRL is undefined and the DC-QAOA
pipeline returns no result whatsoever. The methodology proposed in
Section~\ref{sec:method} is therefore complementary to NRL optimisation: by recovering
validity on graphs where LGP would otherwise abort, FrozenLGP redefines
the regime in which NRL is even a meaningful quantity, while also reducing
the achievable NRL on graphs where standard LGP is forced to use deep,
high-redundancy separator paths to avoid failure.

\section{Graph-Theoretic Foundations of Adaptive Qubit Freezing}
\label{app:graph-theory}

This appendix records the graph-theoretic background for Phase~2
(Sec.~\ref{sec:phase2}). For each already selected candidate $S$, the exact
subproblem is $F=\mathrm{MVC}(G\setminus S)$. Exactness here concerns the
cut cardinality of that residual graph, not joint minimization over $S$ and
$F$, and not the quality of the final variational MaxCut solution.

\subsection{Separators and flow networks}

We use the usual terms: a \emph{graph} $G=(V,E)$ has vertices $V$ (one qubit
each) and edges $E$ (one coupling each); a graph is \emph{connected} when a path
joins every pair of vertices. A set $S\subseteq V$ is a \emph{vertex separator}
(Def.~\ref{def:separator}) if $G\setminus S$ has at least two connected
components.

\begin{definition}[Vertex separator]\label{def:separator}
A set $S\subseteq V$ is a \emph{separator} of $G$ if removing $S$ and its
incident edges leaves $G\setminus S$ with at least two connected components.
\end{definition}

Phase~1 succeeds when such an $S$ exists with $|S|\le k-1$ and yields exactly two
components [Eq.~\eqref{eq:lgp-condition}]. On dense graphs no small separator
exists, so Phase~2 asks a conditional question: for this $S$, what is the
smallest additional vertex set whose removal disconnects $G\setminus S$?
Local source--sink cuts provide the primitive used by the global node-cut
routine. A \emph{flow network} is a directed graph in which every arc
$(u,v)$ has a non-negative \emph{capacity} $c(u,v)$; a \emph{flow} $f$ obeys the
capacity and conservation constraints
\begin{align}
0\;\le\; f(u,v)\;&\le\; c(u,v) && \text{(capacity)},\label{eq:capacity}\\
\sum_{u} f(u,w)\;&=\;\sum_{x} f(w,x) && \forall\, w\neq s,t \;\;\text{(conservation)},
\label{eq:conservation}
\end{align}
and its \emph{value} $|f| = \sum_x f(s,x) - \sum_u f(u,s)$ is the net amount
leaving $s$. The \emph{maximum-flow problem} seeks a feasible flow of largest
value.

\subsection{$s$--$t$ cuts and the minimum cut}

The same network can be viewed from the opposite side, not by what flows
through it, but by where it can be severed.

\begin{definition}[$s$--$t$ cut]\label{def:cut}
An \emph{$s$--$t$ cut} is a partition of the vertices into two sets $(A,B)$ with
$s\in A$ and $t\in B$. Its \emph{capacity} is the total capacity of the arcs
crossing from $A$ to $B$,
\begin{equation}
c(A,B)\;=\;\sum_{\substack{u\in A,\, v\in B \\ (u,v)\in E}} c(u,v).
\label{eq:cutcap}
\end{equation}
\end{definition}

Such a cut is a set of arcs whose removal disconnects $s$ from $t$: once every
$A\!\to\!B$ arc is gone, no material can reach the sink. The
\emph{minimum-cut problem} is to find the cut of smallest capacity. Because
every unit of flow from $s$ to $t$ must cross any given cut at least once, the
value of \emph{any} feasible flow is bounded above by the capacity of
\emph{any} cut: the cut acts as a bottleneck. The narrowest bottleneck
therefore caps the largest flow, a relationship the next theorem makes exact.

\begin{figure*}[t]
\centering
\begin{tikzpicture}[
  every node/.style={font=\small},
  vtx/.style={circle,draw,thick,minimum size=7mm,inner sep=0pt},
  sp/.style={circle,draw,thick,minimum size=8.5mm,inner sep=0pt,fill=blue!6},
]
% (a) an undirected edge in the residual graph G'
\node[vtx] (u) at (0,0) {$u$};
\node[vtx] (v) at (2,0) {$v$};
\draw[thick] (u) -- (v);
\draw[thick,gray!55] (u) -- ++(-1,0.8);
\draw[thick,gray!55] (u) -- ++(-1,-0.8);
\draw[thick,gray!55] (v) -- ++(1,0.8);
\draw[thick,gray!55] (v) -- ++(1,-0.8);
\node at (1,-1.85) {(a) residual graph $G'$};
% mapping arrow
\node at (3.7,0) {\large $\Longrightarrow$};
% (b) node-split directed network
\begin{scope}[xshift=5cm]
  \node[sp] (uin)  at (0,0)   {$u_{\mathrm{in}}$};
  \node[sp] (uout) at (1.9,0) {$u_{\mathrm{out}}$};
  \node[sp] (vin)  at (4.1,0) {$v_{\mathrm{in}}$};
  \node[sp] (vout) at (6.0,0) {$v_{\mathrm{out}}$};
  \draw[->,thick,blue!55!black] (uin) -- node[above=1pt]{$1$} (uout);
  \draw[->,thick,blue!55!black] (vin) -- node[above=1pt]{$1$} (vout);
  \draw[->,thick] (uout) to[bend left=45] node[above=1pt]{$\infty$} (vin);
  \draw[->,thick] (vout) to[out=-90,in=-90,looseness=0.55] node[below=1pt]{$\infty$} (uin);
  \node at (3.0,-1.85) {(b) node-split network $\widehat{G'}$};
\end{scope}
\end{tikzpicture}
\caption{Node-splitting transformation underlying a local minimum-vertex-cut
computation. (a)~An undirected edge $(u,v)$ in the residual graph
$G'=G\setminus S$. (b)~Each vertex is split into an in-copy and an out-copy
joined by a \emph{unit}-capacity arc (blue), while the original edge becomes two
\emph{infinite}-capacity directed arcs. A minimum $s$--$t$ edge cut of
$\widehat{G'}$ can only afford to sever the unit-capacity internal arcs, so its
value equals the size of a minimum vertex set separating the selected
terminals. The global residual cut is obtained by comparing a prescribed
sequence of such local cuts.}
\label{fig:node-split}
\end{figure*}

\subsection{The Maximum-Flow Minimum-Cut theorem}

\begin{theorem}[Max-Flow Min-Cut~\cite{max-min-flow_cut_theorem}]
\label{thm:maxflow}
In any flow network with a single source $s$ and sink $t$, the maximum value of
an $s$--$t$ flow equals the minimum capacity over all $s$--$t$ cuts:
\begin{equation}
\max_{f}\;|f| \;=\; \min_{(A,B)}\; c(A,B).
\label{eq:maxflowmincut}
\end{equation}
\end{theorem}

The bottleneck argument of the previous paragraph already shows that
$\max|f|\le\min c(A,B)$; the content of the theorem is that the two are in fact
\emph{equal}, so the largest flow is throttled by exactly the smallest
bottleneck and nothing else. The result is also constructive. Augmenting-path
algorithms such as Ford--Fulkerson and its breadth-first refinement
Edmonds--Karp~\cite{max-flow-theory, max-min-flow_cut_theorem} build a maximum
flow by repeatedly pushing material along paths that still have spare capacity
in the \emph{residual graph}, the network of leftover, unused capacity. When
no such augmenting path remains, the flow is maximal, and a minimum cut is read
off for free: let $A$ be the set of vertices still reachable from $s$ in the
residual graph and $B=V\setminus A$. Every arc crossing from $A$ to $B$ is then
saturated, and those saturated arcs form a minimum cut. Computing a maximum flow
thus simultaneously \emph{identifies} a minimum cut, the property FrozenLGP
exploits.

\subsection{From edge cuts to the minimum vertex cut}

Theorem~\ref{thm:maxflow} concerns \emph{edge} cuts, but FrozenLGP must freeze
\emph{vertices}: a frozen qubit is a vertex removed from the quantum circuit,
not an edge. We therefore need the smallest set of \emph{vertices} whose
deletion disconnects the residual graph, its \emph{minimum vertex cut} (MVC).
A standard reduction, \emph{node splitting}, recasts this vertex problem as an
edge problem so that the max-flow machinery above applies verbatim
(Fig.~\ref{fig:node-split}).

Given the (undirected) residual graph $G'=(V',E')$, we build a directed
auxiliary network $\widehat{G'}$ as follows:
\begin{itemize}
\item split every vertex $v\in V'$ into an in-copy $v_{\mathrm{in}}$ and an
out-copy $v_{\mathrm{out}}$, joined by an internal arc
$v_{\mathrm{in}}\!\to\!v_{\mathrm{out}}$ of \emph{unit} capacity;
\item replace every edge $(u,v)\in E'$ by two \emph{infinite}-capacity arcs
$u_{\mathrm{out}}\!\to\!v_{\mathrm{in}}$ and
$v_{\mathrm{out}}\!\to\!u_{\mathrm{in}}$.
\end{itemize}
The construction is engineered so that the only arcs a finite-capacity cut can
afford to sever are the unit-capacity \emph{internal} arcs, cutting an
infinite-capacity edge arc would cost infinitely much. But severing the internal
arc $v_{\mathrm{in}}\!\to\!v_{\mathrm{out}}$ is exactly the act of deleting
vertex $v$ from $G'$. A minimum edge cut of $\widehat{G'}$ between a source $s$
and a sink $t$ therefore selects a set of unit arcs of minimum total capacity,
i.e.\ a minimum-cardinality set of \emph{vertices} whose removal separates $s$
from $t$ in $G'$, and the numerical value of the cut equals the number of those
vertices.

This correspondence is the combinatorial face of a classical result.

\begin{theorem}[Menger~\cite{dittmann2017menger}]\label{thm:menger}
For two non-adjacent vertices $s,t$ of a graph, the minimum number of vertices
whose removal disconnects $s$ from $t$ equals the maximum number of pairwise
internally vertex-disjoint $s$--$t$ paths.
\end{theorem}

Menger's theorem is the vertex analogue of Theorem~\ref{thm:maxflow}, and the
unit-capacity internal arcs are precisely what turn ``number of disjoint paths''
into a flow value: each vertex can relay at most one unit, so the maximum flow
counts vertex-disjoint routes while the minimum cut counts the vertices that
block them. A local node-split computation therefore returns a minimum
terminal-separating set; the global routine compares sufficient terminal
pairs to obtain a minimum cut of the fixed residual graph.

\begin{algorithm}[t]
\caption{Global minimum-node-cut routine used in Phase~2 (NetworkX 3.6.1~\cite{hagberg2008exploring}).}
\label{alg:global-min-node-cut}
\begin{algorithmic}[1]
\Require Undirected residual graph $G'=(V',E')$.
\Ensure A minimum-cardinality vertex cut $F$.
\If{$|V'|<2$ \textbf{or} $G'$ is disconnected}
  \State \textbf{raise input error}
\EndIf
\State $H\gets\textsc{NodeSplitAuxiliary}(G')$
\State $R\gets\textsc{ResidualNetwork}(H)$
\State $v\gets$ first minimum-degree vertex in the iteration order of $G'$
\State $F\gets N_{G'}(v)$ \Comment{initial incumbent; complete-graph case}
\ForAll{$w\in V'\setminus\bigl(N_{G'}(v)\cup\{v\}\bigr)$}
  \State $C\gets\textsc{MinimumSTNodeCut}(G',v,w;H,R)$
  \If{$|C|\leq |F|$}
    \State $F\gets C$
  \EndIf
\EndFor
\ForAll{unordered pairs $\{x,y\}\subseteq N_{G'}(v)$}
  \If{$(x,y)\notin E'$}
    \State $C\gets\textsc{MinimumSTNodeCut}(G',x,y;H,R)$
    \If{$|C|\leq |F|$}
      \State $F\gets C$
    \EndIf
  \EndIf
\EndFor
\State \Return $F$
\end{algorithmic}
\end{algorithm}

Algorithm~\ref{alg:global-min-node-cut} reproduces the control flow used in
the experiments rather than substituting an all-pairs abstraction. The local
\textsc{MinimumSTNodeCut} primitive protects its two terminals from removal as
implemented: in the auxiliary graph it selects the source out-copy
\texttt{mapping[s]B} and target in-copy \texttt{mapping[t]A}, then removes
$\{s,t\}$ from the recovered original-node cut. It reuses the shared node-split
auxiliary graph $H$ and residual network $R$ and, with
\texttt{flow\_func=None}, invokes the pinned default Edmonds--Karp flow solver.
The non-strict update deliberately replaces the incumbent on an
equal-cardinality cut, matching the library implementation. For the selected
reference vertex $v$, the number of local flow calls is
\begin{equation}
q(G',v)=|V'|-1-d(v)
+\left|\left\{\{x,y\}\subseteq N_{G'}(v):(x,y)\notin E'\right\}\right|,
\label{eq:mvc-flow-count}
\end{equation}
which is $\mathcal O(|V'|^2)$ in the worst case. On a complete graph both
loops are empty and the routine returns the initialized cut
$N_{G'}(v)=V'\setminus\{v\}$.

\begin{proposition}[Fixed-residual minimum-cut guarantee]
\label{prop:fixed-residual-mvc}
Let $G'$ be a connected undirected graph with at least two vertices. If $G'$ is
not complete, Algorithm~\ref{alg:global-min-node-cut} returns a set $F$ such
that $G'\setminus F$ is disconnected and $|F|=\kappa(G')$, the vertex
connectivity of $G'$. If $G'=K_r$, it returns $r-1=\kappa(K_r)$ vertices; the
outer Phase~2 feasibility check rejects that result because only one active
vertex remains. The proposition applies to this fixed $G'=G\setminus S$ and
does not imply minimization of $|F|$ over the outer separator choices $S$.
\end{proposition}

\emph{Proof sketch.}
For every non-adjacent terminal pair tested by the global routine, node
splitting and the max-flow/min-cut theorem make the returned local cut a
minimum-cardinality terminal-separating vertex set. The minimum-degree
neighbourhood incumbent together with the two terminal-pair loops is the
sufficient-pair construction for global node connectivity; hence their minimum
has cardinality $\kappa(G')$. In a complete graph no non-adjacent pair exists,
so the initialized neighbourhood $V'\setminus\{v\}$ has cardinality $r-1$,
which equals $\kappa(K_r)$. The final statement follows because the outer
algorithm fixes $S$ before calling this routine.

We validated the pinned implementation on the complete population of 995
connected NetworkX Graph Atlas graphs with $2\le |V'|\le7$. In 995/995 cases,
the returned cut cardinality matched both an independent exhaustive
subset-removal search and \texttt{networkx.node\_connectivity}. Targeted
regression tests additionally cover complete graphs, disconnected inputs,
small residuals, and cuts producing more than two components. This validates
the fixed-residual implementation path; it is not evidence of global
optimality over $S$.

\subsection{Application to adaptive qubit freezing}

We can now state precisely what Phase~2 computes. After a candidate separator
$S$ has failed to bipartition $G$, FrozenLGP forms the residual graph
$G' = G\setminus S$ and computes its minimum vertex cut,
\begin{equation}
F \;=\; \mathrm{MVC}\bigl(G\setminus S\bigr),
\label{eq:Fmvc}
\end{equation}
via the node-split construction above. Since $G'$ carries no intrinsic
source--sink pair, the implementation calls
\texttt{minimum\_node\_cut} with $s=t=\texttt{None}$. In NetworkX 3.6.1,
the global algorithm chooses a minimum-degree vertex $v$, initializes the
incumbent with $N(v)$, computes local cuts from $v$ to each non-neighbour,
and also between non-adjacent pairs of neighbours of $v$, while reusing the
auxiliary and residual networks. It therefore uses a number of flow calls
given exactly by Eq.~\eqref{eq:mvc-flow-count}, with quadratic worst-case
growth. The returned $F$ is a
minimum-cardinality set whose removal disconnects this fixed residual into
at least two components.

The decision rule is then a single comparison against the freeze budget. If
\begin{equation}
|F| \;\le\; m_{\max},
\label{eq:budgetcheck}
\end{equation}
the obstruction is cheap enough to remove: the vertices in $F$ are frozen, their
couplings are folded into the linear bias terms of their surviving neighbours
(\ref{sec:hbias}), and the now-disconnected residual graph yields the valid
bipartition $(G_1,G_2,F)$ returned by Phase~2. Should the minimum cut leave more
than two components, they are merged into exactly two groups, the largest
component versus the union of the rest, so that the downstream two-way
reconstruction applies unchanged. If instead $|F|>m_{\max}$ for every admissible
separator, no affordable freeze set is accepted and control passes to the
coverage-preserving fallback of Phase~3 (Sec.~\ref{sec:cpp}).

Equation~\eqref{eq:Fmvc} freezes the fewest vertices needed to disconnect
the residual for the current $S$. Ties may yield different cut sets with the
same cardinality, and the outer search returns the first budget-feasible
separator rather than the globally smallest $F$ over all $S$. Minimizing
$|F|$ bounds the number of conditional variables and the per-partition
assignment factor; it does not minimize field magnitude, which also depends
on incident weights, degrees, and selected spins.

\section{Computational Overhead and Complexity}
\label{sec:overhead}

FrozenLGP introduces classical partition work, repeated conditional quantum
solves, and reconstruction work. The freeze budget bounds the assignment
factor of one partition call, but it does not by itself bound total recursive
work.

\emph{Classical partition search.} Phase~1 examines at most
$Q_k(n)=\sum_{s=1}^{k-1}\binom{n}{s}$ separator candidates. Phase~2 may
examine the same number again and, for every connected residual, invokes a
global node-cut routine requiring $\mathcal O(n^2)$ local flow calls in the
worst case. If
$T_{\mathrm{flow}}(n,|E|)$ denotes the selected maximum-flow cost, a
conservative expression is
$\mathcal O(Q_k(n)n^2T_{\mathrm{flow}}(n,|E|))$. This is polynomial only
when hardware capacity $k$ is treated as fixed and is too costly for the
$n=10^4$ study; that study uses the explicitly separate scalable upper-bound
driver in \ref{app:driver-agreement}.

\emph{Recursive quantum workload.} At an internal decomposition-tree node
$v$, let $a_v=1$ for Phase~1 or CPP, $a_v=2^{|F_v|-1}$ when the bit-flip
guard applies, and $a_v=2^{|F_v|}$ otherwise. Although
$a_v\le2^{m_{\max}}$, the number of executed leaf solves follows
\begin{equation}
W(v)=a_v\bigl(W(c_1)+W(c_2)\bigr),\qquad W(\text{leaf})=1,
\label{eq:recursive-work}
\end{equation}
not the additive structural leaf count. Assignment factors therefore
multiply along a root-to-leaf path. The recursive-work audit of
Table~\ref{tab:recursive-work} cross-checks this recurrence
against expanded execution and quantifies the resulting inflation.

\emph{Classical MDR and CPP rescoring.} The reconstruction step processes
$|D_1|\cdot|D_2|$ candidate bitstring pairs per assignment, which after
truncation to top-$t$ states ($t=20$ in our experiments) reduces to
$\mathcal{O}(t^2)$ operations per merge and assignment. Direct CPP ranking
adds $\mathcal{O}(|E|t)$ classical edge tests for the retained candidates.
Every individual sub-circuit respects width $k$, and freezing removes
two-qubit terms incident on frozen vertices while adding only local fields.
This can reduce two-qubit gates \emph{per circuit}, as in
Fig.~\ref{fig:freeze_cross_node}. 

In summary, the local assignment factor is bounded, whereas total recursive
work is data dependent and potentially multiplicative. CPP adds no new
quantum term for omitted cross edges but does add classical ranking and can
change the quantum-handled fraction of the objective. The evaluation reports
these costs separately.

\section{Augmented Sub-Circuit Hamiltonian and Bit-Flip Symmetry}
\label{sec:hbias}

Let $(G_1, G_2, F)$ be the output of FrozenLGP, and consider a single
sub-graph $G_S=(V_S,E_S)$ to be solved on quantum hardware. Each frozen
vertex $v\in F$ is assigned a classical Ising spin
$s_v\in\{+1,-1\}$; collecting all such assignments yields a frozen
configuration $\mathbf{s}_F=(s_v)_{v\in F}\in\{+1,-1\}^{|F|}$. Following the
folding identity derived in Section~\ref{sec:prelim_qubit_freezing}, every edge $(v,u)\in E$ with
$v\in F$ and $u\in V_S$ converts a two-body coupling
$J_{v,u}Z_v Z_u$ into a single-body linear term $J_{v,u}s_v Z_u$, and the
augmented cost Hamiltonian for the sub-circuit becomes
\begin{equation}
\label{eq:Haug}
H_S^{\text{aug}}(\mathbf{s}_F)
\;=\;\sum_{(u,v)\in E_S} J_{u,v}\, Z_u Z_v
\;+\;\sum_{u\in V_S} h_u(\mathbf{s}_F)\, Z_u ,
\end{equation}
where the effective linear bias collected by qubit $u\in V_S$ is
\begin{equation}
\label{eq:hbias}
h_u(\mathbf{s}_F)
\;=\;\sum_{\substack{v\in F\\(v,u)\in E}} J_{v,u}\, s_v.
\end{equation}
The first sum in Eq.~\eqref{eq:Haug} retains active-edge couplings and the
second is the conditional influence of $F$. In the demonstrated unweighted
MaxCut setting, $J_{u,v}=w_{u,v}/2=1/2$. We construct a sparse Pauli operator
containing $Z_uZ_v$ and $Z_u$ generators. Under the Qiskit half-angle
convention, their circuit angles are $2\gamma_\ell J_{u,v}$ and
$2\gamma_\ell h_u$, respectively (Sec.~\ref{sec:prelim_qaoa_mapping}).

Equation~\eqref{eq:Haug} is conditional on
$\mathbf{s}_F\in\{+1,-1\}^{|F|}$, giving up to $2^{|F|}$ subproblems. For
any weighted or unweighted edge-only MaxCut Hamiltonian with zero fields,
the bit-flip symmetry
$C(\mathbf{s})=C(-\mathbf{s})$ recalled in Sec.~\ref{sec:prelim_qubit_freezing} implies that
the configurations $\mathbf{s}_F$ and $-\mathbf{s}_F$ produce identical cut
values and identical (up to a global bit-flip) bitstring distributions. We
therefore fix the last frozen vertex and
enumerate only the $2^{|F|-1}$ remaining configurations on hardware. This
reduces the assignment count from $2^{|F|}$ to $2^{|F|-1}$ at that
partition without changing the paired conditional spectra.

% \paragraph{Validity of the halving under recursion.}
The argument above requires the active-qubit field to vanish: the joint
flip $\mathbf{s}_F\!\to\!-\mathbf{s}_F$ and $\mathbf{z}\!\to\!-\mathbf{z}$
leaves the couplings $J_{u,v}Z_uZ_v$ invariant and flips the
\emph{frozen-induced} bias $h_u(\mathbf{s}_F)$ of Eq.~\eqref{eq:hbias}, so the
two spectra coincide. When FrozenLGP is invoked recursively, however, a child
sub-graph may inherit a nonzero external bias $\{h_u^{\text{ext}}\}$ propagated
from its parent (~\ref{sec:mdr}). Because $h_u^{\text{ext}}$
does \emph{not} change sign under $\mathbf{s}_F\!\to\!-\mathbf{s}_F$, the term
$\sum_u h_u^{\text{ext}}Z_u$ tilts the landscape and breaks
$C(\mathbf{z})=C(-\mathbf{z})$; the factor-of-two saving is then no longer
exact. The orchestrator therefore applies the halving \emph{only} at levels
where the inherited active-qubit bias is identically zero, and enumerates all
$2^{|F|}$ assignments otherwise. Although every \emph{initial} field is zero in
the unweighted MaxCut benchmark of Sec.~\ref{sec:evaluation}, recursion does
propagate nonzero $h^{\text{ext}}$ to child sub-graphs that themselves freeze,
so the guarded full enumeration is exercised at those levels.

% \paragraph{Weighted MaxCut and general QUBO.}
Weighted MaxCut with edge terms only retains the same global bit-flip
symmetry, so unequal positive edge weights do not disable halving. What
disables it is a field term that does not flip with the frozen assignment:
for example, an initial linear term in a general Ising/QUBO objective or a
bias inherited from a parent FrozenLGP call. Then all $2^{|F|}$ assignments
are enumerated. The freeze budget still bounds this \emph{local} factor by
$2^{m_{\max}}$, but Eq.~\eqref{eq:recursive-work} shows why the total
recursive workload is not independent of $n$. The present experiments test
unweighted MaxCut; broader Ising objectives are supported algebraically but are not empirically claimed here.

\section{Measurement Distribution Reconstruction of FrozenLGP}
\label{sec:mdr}

The MDR step merges sub-circuit outputs by enforcing
agreement on the bits assigned to separator nodes shared by $G_1$ and
$G_2$. FrozenLGP must additionally handle the bits associated with the
frozen set $F$, which are \emph{classical constants} rather than quantum
measurement outcomes. We therefore extend MDR with three modifications,
formalised in Algorithm~\ref{alg:mdr}.

First, the separator-compatibility check is restricted to the genuinely
shared quantum nodes $S=V_{G_1}\cap V_{G_2}$: a pair of sub-distribution
samples $(b_1,b_2)$ is admissible if and only if their bits agree at every
index corresponding to a node of $S$. Frozen vertices are absent from both
sub-graphs by construction and therefore make no contribution to the
compatibility predicate. Second, after a compatible pair has been
identified, the parent bitstring is assembled in the recorded order
$\mathcal{O}=\bigl[V_{G_1},\,V_{G_2}\setminus V_{G_1},\,F\bigr]$.
At a leaf, bit position $i$ maps to the $i$-th vertex in the subgraph's
recorded insertion order; the
backend reverses Qiskit's little-endian count strings once to enforce that
convention. A merge preserves the first child's order, appends
second-child-only labels in their recorded order, and then appends the
sorted frozen labels. Every raw benchmark record stores this original-label
node ordering. Third, the frozen positions are filled
deterministically from the current spin assignment $\mathbf{s}_F$ using
the canonical mapping $z_v=(1-s_v)/2$, i.e. a bit value of $0$ encodes
spin $+1$ and a bit value of $1$ encodes spin $-1$.

For the scoring step we adopt the \emph{minXmul} scheme of the underlying
DC-QAOA framework~\cite{dc-qaoa}, which assigns each reconstructed
bitstring the weight
\begin{equation}
\label{eq:minxmul}
\text{score}(\mathbf{z})\;=\;\min(c_1,c_2)\cdot c_1\cdot c_2 ,
\end{equation}
where $c_1$ and $c_2$ are the sample counts of the two contributing
sub-circuit bitstrings $b_1$ and $b_2$. This product-with-minimum
formulation favours high-probability sub-circuit modes while penalising
pairs whose sub-graph support is unbalanced. The reconstructed distribution
is finally sorted, truncated to the top $t$ states, and rescaled to the
global shot budget. Consequently, its distributional metrics are conditional
on the retained reconstruction set rather than estimates of a complete joint
distribution.

\begin{algorithm}[t]
\caption{Extended MDR with frozen constants.}
\label{alg:mdr}
\begin{algorithmic}[1]
\Require Sub-distributions $D_1,D_2$ over node orderings $V_{G_1},V_{G_2}$;
         shared separator $S=V_{G_1}\cap V_{G_2}$; frozen set $F$ with
         spin assignment $\mathbf{s}_F$; scoring scheme $\phi$.
\Ensure  Parent distribution $D_P$ over the ordered node set
         $\mathcal{O}$.
\State $\mathcal{O}\gets\bigl[V_{G_1},\,V_{G_2}\setminus V_{G_1},\,F\bigr]$
\State $D_P\gets\emptyset$
\ForAll{$(b_1,c_1)\in D_1,\,(b_2,c_2)\in D_2$}
  \If{$b_1[i]=b_2[j]$ for every separator index pair $(i,j)\in I_S$}
    \State $b_{\text{Q}}\gets b_1\,\Vert\,b_2\!\restriction_{V_{G_2}\setminus V_{G_1}}$
    \State $b_{\text{F}}\gets\bigl((1-s_v)/2\bigr)_{v\in F}$
    \State $b\gets b_{\text{Q}}\,\Vert\,b_{\text{F}}$
    \State $D_P[b]\mathrel{+}{=}\phi(c_1,c_2)$
  \EndIf
\EndFor
\State sort $D_P$, truncate to top-$t$, rescale to shot budget
\State \Return $(D_P,\mathcal{O})$
\end{algorithmic}
\end{algorithm}

When FrozenLGP is invoked recursively, the bias contributions propagated
by each level of decomposition compose linearly: a sub-graph that itself
inherits external biases $\{h_u^{\text{ext}}\}$ from a parent call simply
adds them to the freshly computed bias of Eq.~\eqref{eq:hbias} before
constructing $H_S^{\text{aug}}$. The orchestrator manages this propagation transparently, repeating
the bias-enumerate-solve-reconstruct loop at every recursion level until
the leaf sub-graphs fit within the $k$-qubit budget.

\section{Analyze Graph Partition Scalability}\label{app:extend_partition_result}

\begin{figure*}[t]
    \centering
    \includegraphics[width=\textwidth]{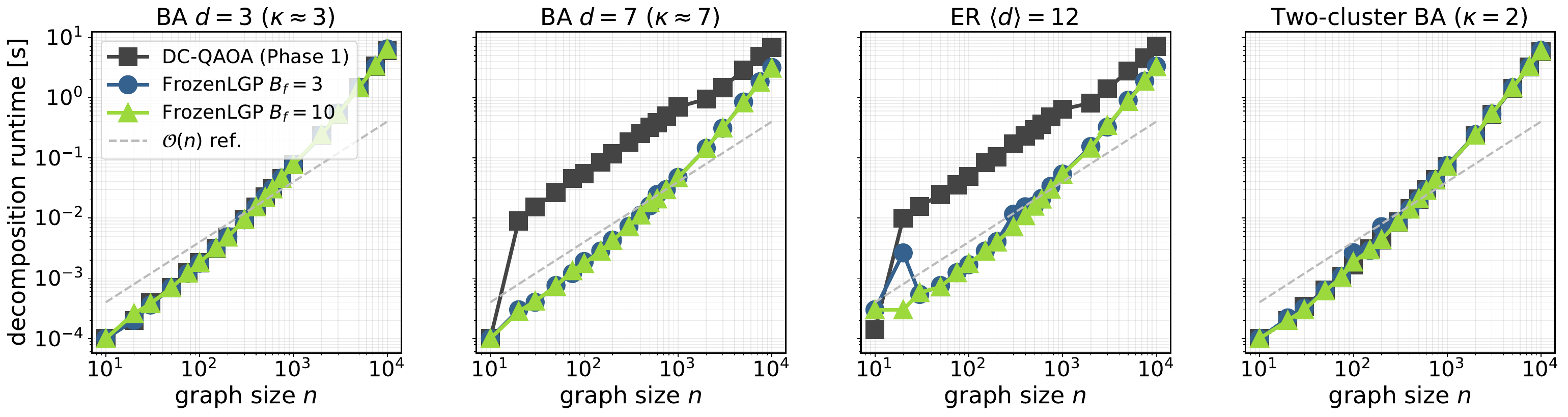}
    \caption{Full-decomposition runtime versus graph size for the four
    benchmark families (log--log; means over seeds). FrozenLGP
    ($B_f{=}3$ and $B_f{=}10$) and the Phase-1-only DC-QAOA baseline
    scale near-linearly (dashed $\mathcal{O}(n)$ reference). On the
    families whose connectivity exceeds the separator budget (BA $d{=}7$,
    dense ER), the baseline pays a persistent overhead for repeatedly
    probing uncuttable sub-problems before failing, whereas FrozenLGP
    removes the obstructing vertices and proceeds; at $n=10^4$ this
    yields a ${\approx}2.1\times$ runtime advantage (3.2\,s vs.\ 6.8\,s
    on BA $d{=}7$; 3.3\,s vs.\ 7.1\,s on dense ER). On the
    $\kappa \le k-1$ graph families the two methods coincide.}
    \label{fig:scal-runtime}
\end{figure*}

\subsection{Runtime and memory scaling}
Fig.~\ref{fig:scal-runtime} shows that full decomposition cost grows
near-linearly with $n$ for both methods and all families: an
$n=10{,}000$ instance is fully decomposed in 3--7\,s on a single core.
Two method-level differences emerge. First, on the high-connectivity
families FrozenLGP is consistently \emph{faster} than the baseline
(${\approx}2.1\times$ at $n=10^4$), because the baseline expends
max-flow probes on sub-problems that admit no $(k{-}1)$-separator before
aborting, whereas freezing eliminates the obstruction and continues
peeling cheaply. Second, the same effect dominates memory: at $n=1000$
the baseline's peak traced allocation averages 16.2\,MB (flow-network
construction on uncuttable sub-problems) versus 0.9\,MB for FrozenLGP.
These measurements show that the scalable \emph{preprocessing driver} is not
the bottleneck in this CPU benchmark. They do not include conditional circuit
optimisation, sampling, or reconstruction and therefore do not establish that
the complete FrozenLGP pipeline is cheaper.

\begin{figure*}[t]
    \centering
    \includegraphics[width=\textwidth]{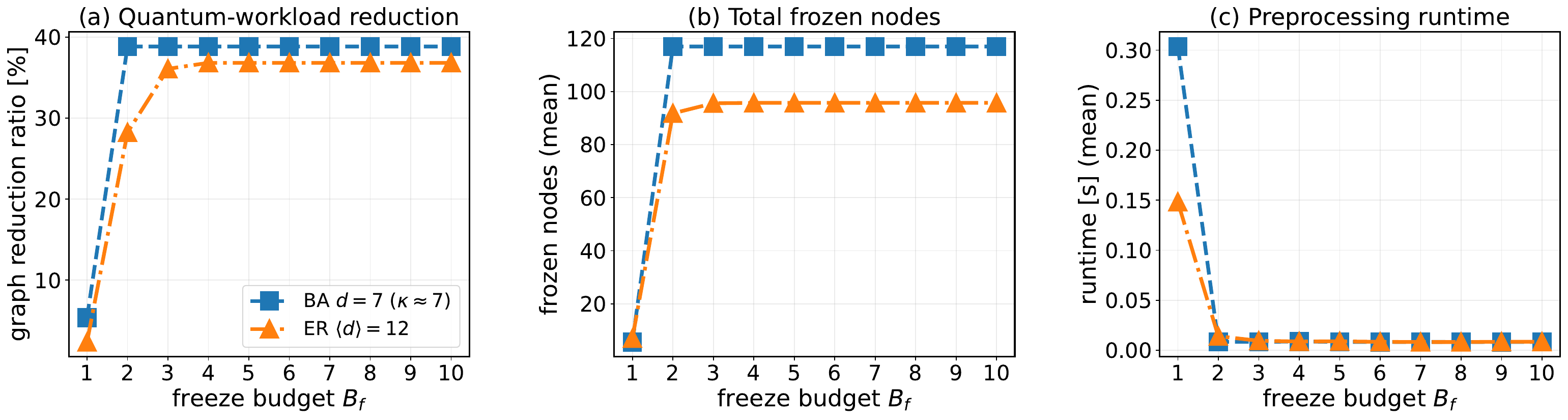}
    \caption{Budget dependence on the high-connectivity families
    ($n\le1000$ pooled). (a)~Graph reduction ratio $\rho$ and (b)~total
    frozen nodes change little beyond the observed coverage knee. The driver
    accepts the first feasible cut within the budget rather than necessarily
    consuming the full budget. (c)~Preprocessing runtime is flat beyond the
    observed knee; the elevated cost at $B_f{=}1$
    reflects max-flow probing on instances that ultimately require CPP.}
    \label{fig:scal-budget}
\end{figure*}

\subsection{Graph reduction}\label{subsec:graph_reduction}
Fig.~\ref{fig:scal-budget} quantifies the freezing activity itself. At
and beyond the budget threshold, the cumulative reduction ratio
saturates at $\rho \approx 38$--$39\%$ on BA $d{=}7$ and
$\rho \approx 36$--$37\%$ on dense ER, independent of both $B_f$ and $n$, per peel step the algorithm freezes approximately
$\kappa - (k-1)$ vertices wherever the local cut exceeds the separator
budget, so the aggregate fraction is empirically stable on these samples. Each frozen
vertex leaves the quantum register and is folded into the
$h$-bias of its active neighbours (Sec.~\ref{sec:prelim_qubit_freezing}), so on dense
instances FrozenLGP removes over a third of the active graph vertices in the
structural decomposition accounting. This can reduce individual circuit width
and edge count, but it is not a reduction in aggregate execution work. Equally
important is the converse: on the
$\kappa \le k-1$ families the frozen count is \emph{exactly zero} in all
1{,}180 runs, and the decomposition trees produced by FrozenLGP and the
baseline are identical instance-by-instance, an at-scale empirical
confirmation of the strict phase-priority guarantee that FrozenLGP never
incurs freezing overhead where standard LGP suffices.

\begin{figure*}[t]
    \centering
    \includegraphics[width=\textwidth]{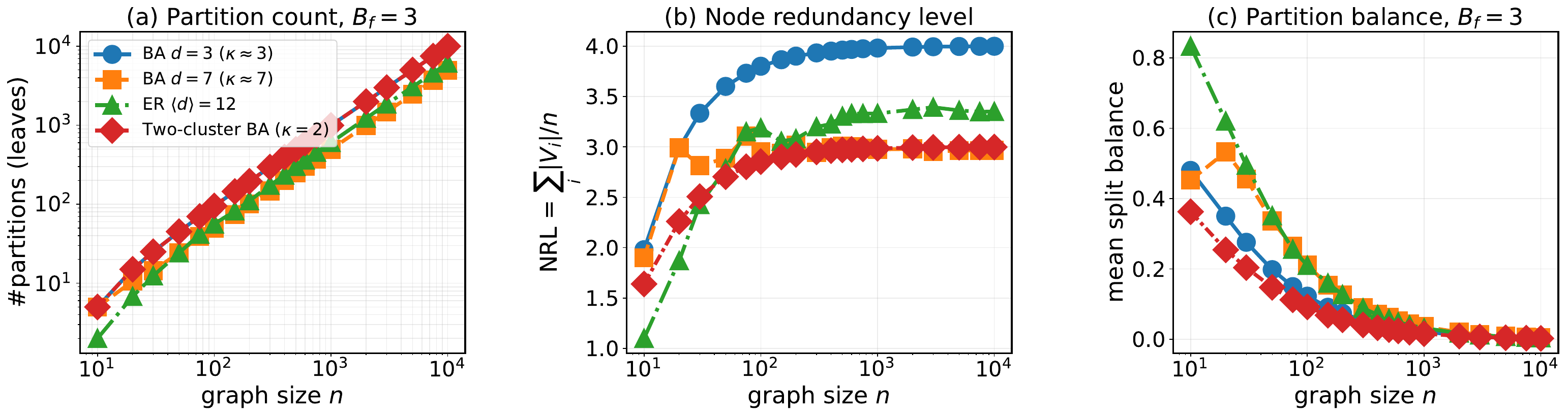}
    \caption{Partition quality versus graph size for FrozenLGP
    ($B_f{=}3$). (a)~The number of leaf sub-problems grows linearly with
    $n$. (b)~The node redundancy level saturates at $3.0$--$4.0$,
    reflecting separator duplication across the decomposition; a larger
    freeze budget lowers structural NRL since frozen vertices leave the leaf
    graphs (on instances the baseline can also decompose the two NRLs
    coincide by the phase-priority guarantee, so this is a within-FrozenLGP
    effect rather than a reduction below the baseline). (c)~Mean split balance decays
    with $n$: minimum-cut separators preferentially peel small fringe
    components, a structural property of vertex-cut partitioning shared
    by the baseline.}
    \label{fig:scal-quality}
\end{figure*}

\subsection{Partition quality and structural limitations}\label{subsec:partition_quality_structure}
Fig.~\ref{fig:scal-quality} examines the decompositions themselves.
The number of leaf sub-problems grows linearly in $n$ (0.5--1.0
leaves per vertex at $B_f{=}3$) with leaf sizes of 3--6 vertices, and
NRL saturates at 3.0--4.0 across families: each original vertex is
submitted to the QAOA solver roughly three to four times due to
separator duplication. Both properties transfer unchanged from standard
LGP, on the easy families the baseline produces identical values,
and freezing reduces NRL slightly on dense graphs (3.33 vs.\ 3.50 at the
extreme scale) by excluding frozen vertices from both children. The
principal structural limitation is split balance: because minimum
vertex cuts of sparse-to-moderate random graphs overwhelmingly isolate
small fringe components, the mean balance $\beta$ decays from
${\approx}0.32$ at small scale to below $0.01$ at $n=10^4$, and the
decomposition depth grows linearly rather than logarithmically. This is
intrinsic to minimum-cut-driven vertex-removal partitioning~\cite{graph_partition_irregular_graph, dittmann2017menger, max-flow-theory, graph_theory_west2001introduction} rather than
to the freezing mechanism, the baseline exhibits the same decay
wherever it succeeds, and the CPP fallback, whose median split is
balanced by construction, is unaffected. The practical consequence is a
long sequential chain of small sub-problems rather than a shallow
parallel tree; incorporating balanced-separator search into the
candidate generation is a natural extension, which we leave to future
work. Moreover, a structural leaf is not a single execution when multiple
recursive nodes freeze variables. The recursive-work audit in
Table~\ref{tab:recursive-work} finds multiplicative assignment inflation as
large as $1.17\times10^6$ on one BA $d=7$ instance. The linear leaf-count and
NRL curves must therefore not be read as total quantum-work curves.

\begin{figure}[t]
    \centering
    \includegraphics[width=0.8\columnwidth]{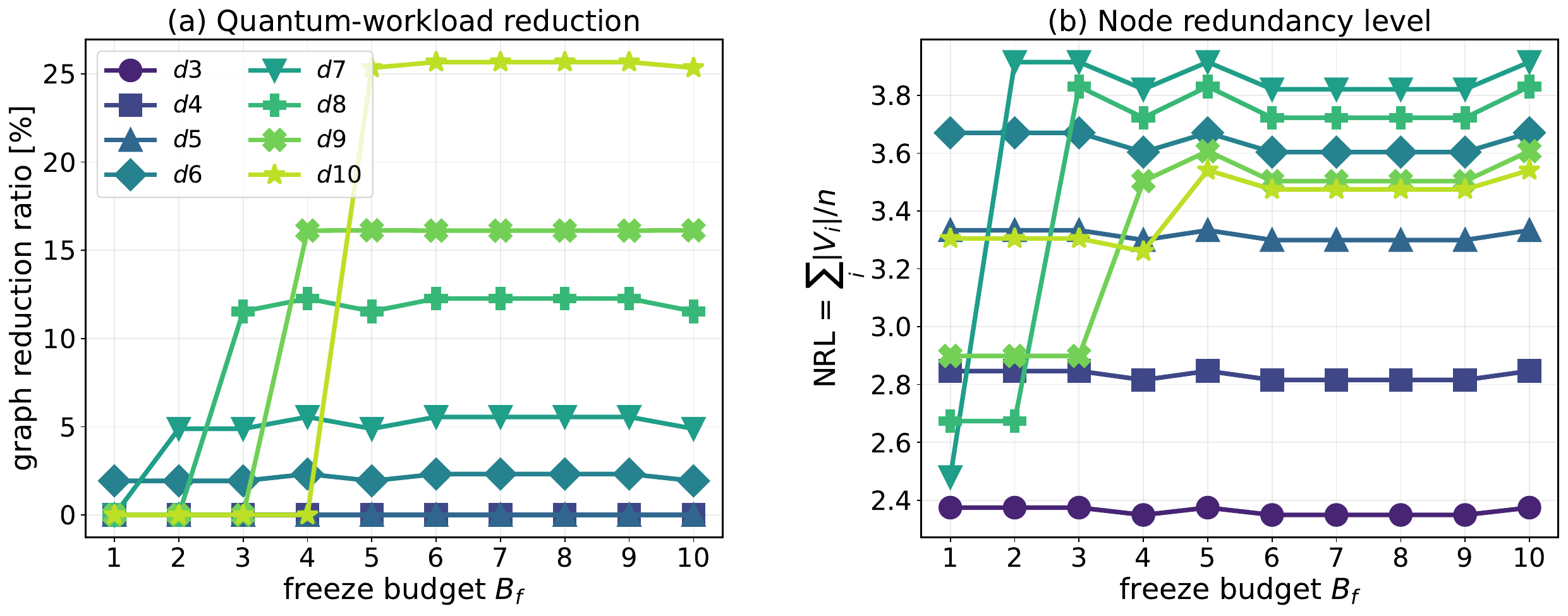}
    \caption{Budget dependence on random $d$-regular graphs
    ($n\le1000$, successful runs). (a)~Graph reduction ratio $\rho$ and
    (b)~node redundancy level, per degree. Beyond the empirical staircase
    $B_f^{*}=d-(k-1)$ both quantities are nearly budget-independent in this
    sample; the
    reduction grows with the connectivity surplus, from ${\approx}1$--5\%
    at $d{=}7$ to ${\approx}25\%$ at $d{=}10$, while NRL peaks near
    $d{=}8$ (${\approx}4.3$) and decreases again at higher degrees as
    freezing removes a larger fraction of vertices from the structural leaf
    graphs.}
    \label{fig:reg-reduction}
\end{figure}

\subsection{Reduction, redundancy, and runtime of Regular Graph}\label{subsec:reduce_redun_runtime_regular}
Beyond the observed staircase, the freezing fraction changes little with the
allowed budget (Fig.~\ref{fig:reg-reduction}) and grows with the sampled
connectivity surplus
$d-(k-1)$, from ${\approx}1$--5\% at $d{=}7$ to ${\approx}9\%$ at $d{=}8$,
${\approx}17\%$ at $d{=}9$, and ${\approx}25\%$ at $d{=}10$, i.e.\ on
10-regular graphs one quarter of all vertices is converted into classical
bias terms. The node redundancy level increases from 2.4 at $d{=}3$ to a
maximum of ${\approx}4.3$ at $d{=}8$ and then decreases at higher degrees
as freezing thins the structural leaf graphs. For $d \le 5$ FrozenLGP and the
baseline produce identical decomposition trees with zero frozen vertices,
extending the strict-priority confirmation of~\ref{subsec:graph_reduction} to the regular family. Runtime
scaling remains near-linear across all eight degrees: full decomposition
of $n=10^4$ instances completes in 3.9--6.6\,s, while the baseline spends
4--5\,s probing $d \ge 6$ instances with bounded max-flow before failing;
at $d{=}10$, FrozenLGP completes its decomposition faster
(3.9\,s) than the baseline takes to fail (4.9\,s). Split balance follows
the same fringe-peeling decay as in~\ref{subsec:partition_quality_structure}, confirming that this limitation
is a property of minimum-cut-driven vertex removal rather than of any
particular degree structure~\cite{graph_partition_irregular_graph, dittmann2017menger, graph_theory_west2001introduction}.

\subsection{Exact and polynomial-time partitioners}
\label{app:driver-agreement}

\begin{table}[t]
\caption{\label{tab:driver-agreement}%
Top-level frozen-set cardinality relative to a global-over-$S$ exhaustive
reference on 84 instances from seven families ($n\in\{8,10,12,14\}$,
three seeds, $k=6$, $B_f=5$). A comparison is included only when both outputs
are feasible and satisfy the same connected-group structural condition.
``Relaxed'' counts outputs outside that condition and is reported separately.
The final column gives the number of cases at gaps
$\Delta=0/1/2/3/4$, where $\Delta=|F_{\rm driver}|-|F_{\rm ref}|$.}
\centering
\resizebox{\columnwidth}{!}{%
\begin{tabular}{lrrrrrr}
\hline
Driver & Comparable & $|F|$ equal & Equal [\%] & Mean gap & Max gap &
$n_{\Delta=0/1/2/3/4}$ \\
\hline
Exact production (first feasible $S$) & 69 & 51 & 73.9 & 0.768 & 4 & 51/1/3/10/4 \\
Scalable production & 64 & 64 & 100.0 & 0.000 & 0 & 64/0/0/0/0 \\
Large-scale figure driver & 63 & 63 & 100.0 & 0.000 & 0$^{\dagger}$ & 63/0/0/0/0 \\
\hline
\end{tabular}
}
\par\smallskip
\raggedright\footnotesize $^{\dagger}$Six additional outputs used relaxed,
not-necessarily-connected groupings and are excluded from the cardinality gap.
\end{table}

The exhaustive separator search of Algorithm~\ref{alg:frozenlgp} is
$\mathcal{O}(n^{k-1})$ and hence intractable at the scales studied in
Sec.~\ref{subsec:partition-scalability}. The partition-level results there are
therefore produced by a polynomial-time upper-bound driver. It selects cuts
from minimum-degree neighbourhoods and budget-bounded source--sink flow probes,
then peels subproblems in place. These feasible cuts are not certificates of a
globally minimal frozen set.

To quantify the distinction, the driver-agreement audit enumerates every
admissible $S$ on
the tractable set and minimizes $|F|$ over all budget-feasible residual cuts.
The production exact procedure returns the first feasible $S$ by design;
it matches this stronger reference in 51 of 69 comparable cases, with mean
cardinality gap 0.768 and maximum gap four. Its full observed gap distribution
is 51/1/3/10/4 cases at $\Delta=0/1/2/3/4$, respectively. This nonzero tail
arises because the production procedure stops at the first budget-feasible
separator rather than scoring all admissible $S$. The exactness guarantee is thus
conditional on the selected $S$. The scalable and figure drivers
happen to have zero gap on their 64 and 63 structurally comparable audit cases,
respectively, but this finite observation is not a proof. All reference searches
completed without timeout. Per-instance graph fingerprints, selected $S$ and
$F$, group-connectivity flags, and gaps are included in the reproducibility
artifacts.

\begin{table*}[t]
\caption{\label{tab:noise-evs}%
Paired noise study on 60 exact-ground-truth instances
($n\in\{8,10,12\}$; 360 records). AR is the best sampled candidate divided by
the optimum; EVS is the mean cut value of the reported distribution divided by
the optimum. $P_{\rm opt}$ and $P_{\ge.95}$ are its mass at the optimum and at
AR$\ge0.95$. For direct QAOA these are full sampled-distribution metrics; for
DC-QAOA and FrozenLGP they are conditional on the top-20 reconstruction retained
by the algorithm. Best AR is reported as the mean, median, and worst solved-case
value; the distribution metrics are means over solved instances. CVaR is not
reported because a tail-risk comparison between complete monolithic counts and
conditional top-20 reconstructed support would conflate algorithmic truncation
with physical-noise response.}
\centering
\resizebox{\textwidth}{!}{%
\begin{tabular}{llrrrrrrr}
\hline
Method & Noise & Coverage & Best AR mean & Best AR median & Best AR worst &
EVS & $P_{\rm opt}$ & $P_{\ge.95}$ \\
\hline
Direct QAOA & depolarizing & 1.000 & 0.962 & 1.000 & 0.667 & 0.757 & 0.042 & 0.049 \\
Direct QAOA & strong & 1.000 & 0.947 & 0.985 & 0.769 & 0.692 & 0.016 & 0.019 \\
DC-QAOA & depolarizing & 0.883 & 0.962 & 1.000 & 0.818 & 0.894 & 0.230 & 0.255 \\
DC-QAOA & strong & 0.883 & 0.973 & 1.000 & 0.857 & 0.855 & 0.190 & 0.209 \\
FrozenLGP ($B_f{=}3$) & depolarizing & 1.000 & 0.964 & 1.000 & 0.818 & 0.896 & 0.215 & 0.288 \\
FrozenLGP ($B_f{=}3$) & strong & 1.000 & 0.975 & 1.000 & 0.846 & 0.862 & 0.175 & 0.196 \\
\hline
\end{tabular}
}
\end{table*}

\begin{figure}[t]
    \centering
    \includegraphics[width=0.8\columnwidth]{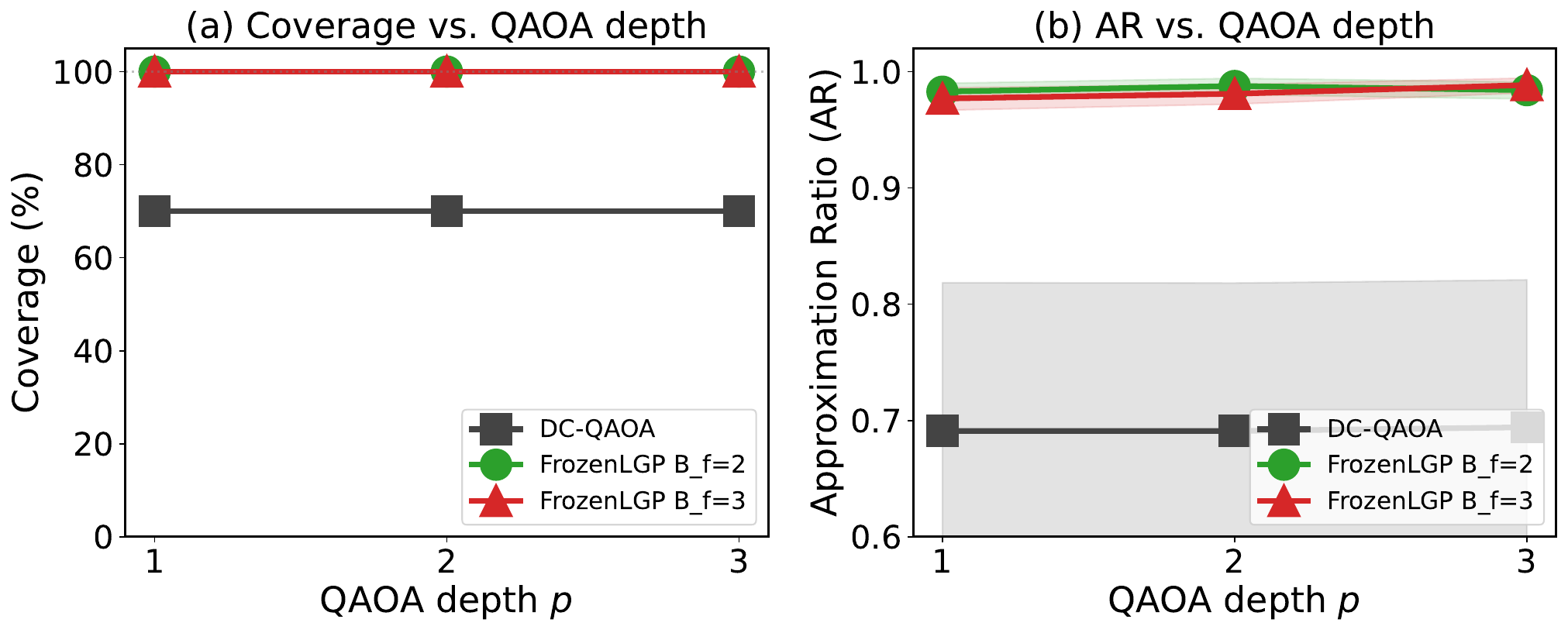}
    \caption{QAOA depth sweep ($p\in\{1,2,3\}$) on 40 dense instances
    (ER $p{=}0.8$ and BA $d{=}3$, $n\in\{8\text{--}16\}$, noiseless).
    (a)~Coverage (\%): DC-QAOA is flat at 70\% at every depth;
    FrozenLGP ($B_f{=}3$) is flat at 100\%.
    (b)~Approximation ratio with bootstrap 95\,\% CI shading.}
    \label{fig:e2e-depth}
\end{figure}

\begin{figure}[t]
    \centering
    \includegraphics[width=0.8\columnwidth]{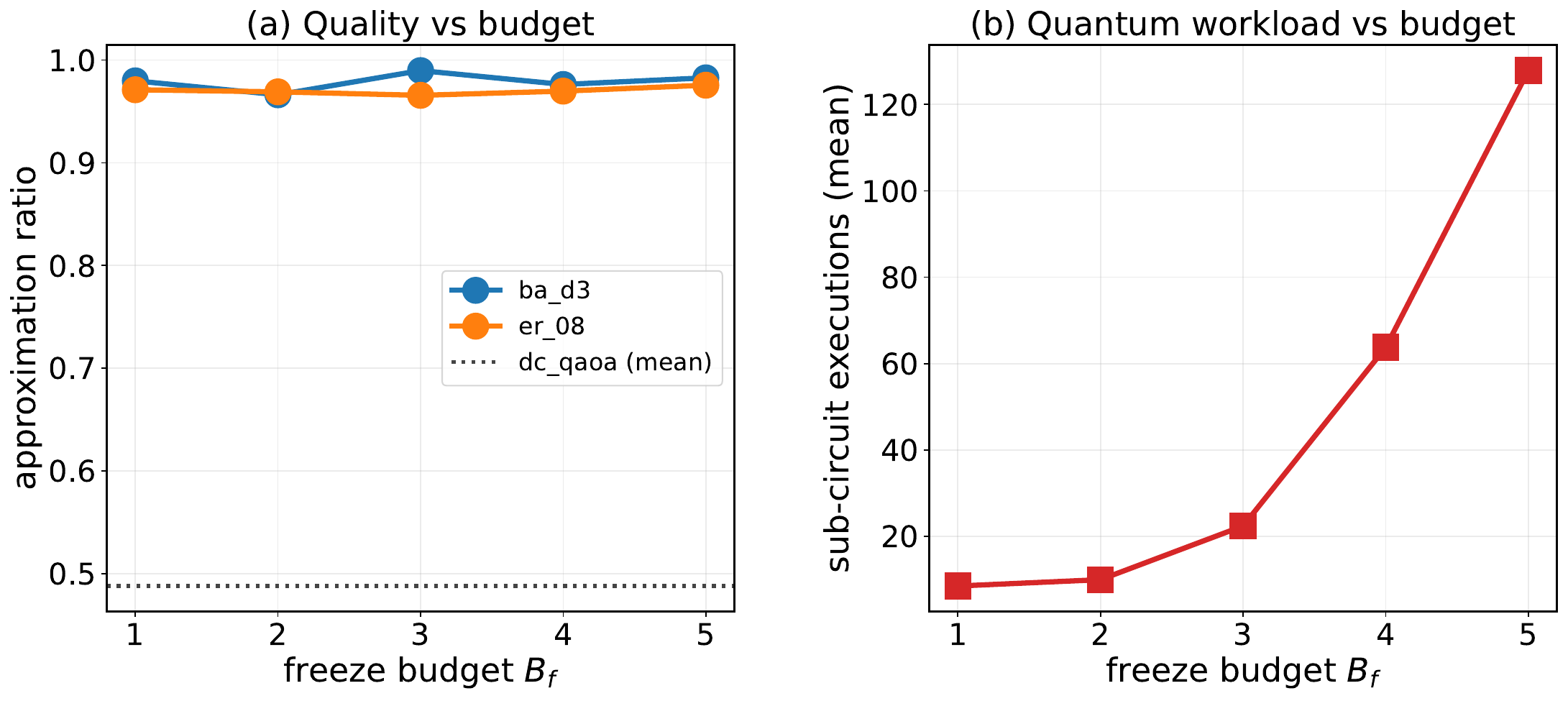}
    \caption{End-to-end budget sweep on 30 hard instances
    (ER $p{=}0.8$ and BA $d{=}3$, $n\in\{12,16,20\}$, noiseless).
    (a)~Approximation ratio per budget (dashed: DC-QAOA at 50\,\%
    coverage on these instances);
    (b)~executed sub-circuit count, reflecting the $2^{|F|}$ assignment
    enumeration.}
    \label{fig:e2e-budget}
\end{figure}

\section{Simulated noise stress test}\label{sec:noise_rubus_validation}
These results are simulations, not executions on physical hardware. We reran
the paired benchmark from raw records using deterministic graph, optimiser,
transpiler, and simulator seeds and retained the optimiser initial point,
graph/circuit fingerprints, node ordering, and final distributions. The common
scope is 60 instances at $n\in\{8,10,12\}$ (four graph families and five seeds),
$p=1$, one deterministic restart, 1024 estimator shots per noisy objective
evaluation, 10{,}000 final shots, and at most 300 COBYLA iterations. Exact
strong-noise simulation of monolithic $n=16$ circuits did not finish within the
validation budget; eight partial decomposed-method records from that size are
archived outside the analysis directory and excluded from every statistic.

All simulations use Qiskit 2.3.1 with Qiskit Aer 0.17.2, pinned in the
released \texttt{environment.yml}. Noiseless benchmarks evaluate analytic
statevector expectation values; under an active noise model, objective
evaluations use the shot-based Aer estimator primitive (1024 shots per
evaluation) and final sampling uses the Aer sampler primitive with the same
noise model and seeds bound at construction. Circuits are transpiled with a
seeded transpiler at optimization level~1 to the
$\{\mathrm{CX},U,R_z,R_y,R_x,H,X\}$ basis on which the noise models define
their errors; no coupling-map constraint is imposed, consistent with the
all-to-all connectivity assumption stated in Sec.~\ref{subsec:discussion-limitations} and ~\ref{subsec:e2e-resources}.

The depolarizing channel uses one-qubit error 0.001, CX error 0.01, and
symmetric readout error 0.01. The strong channel combines $T_1=20\,\mu$s,
$T_2=10\,\mu$s, 50-ns one-qubit and 400-ns two-qubit durations, one-qubit
depolarizing error 0.02, CX error 0.05, and readout error 0.05. The same channel
is active during the shot-based objective evaluations and final sampling.
These are declared stress models, not device-calibrated models.

Table~\ref{tab:noise-evs} separates best-candidate AR from distributional
quality. Under the stronger channel the mean EVS decreases from 0.757 to 0.692
for direct QAOA, from 0.894 to 0.855 for DC-QAOA, and from 0.896 to 0.862 for
FrozenLGP. The mean probability of sampling an optimum also decreases in all
three cases. Across solved instances, the median best AR is 1.000 except for
direct QAOA under strong noise (0.985), whereas the worst best AR ranges from
0.667 to 0.857; these tails reinforce that the mean alone is insufficient.
Best-sampled AR is not monotone for the decomposed methods
(0.962 to 0.973 for DC-QAOA and 0.964 to 0.975 for FrozenLGP), illustrating why
best-of-10{,}000 post-selection alone is not a reliable noise-robustness metric.
DC-QAOA has 53/60 operational coverage under both channels; FrozenLGP and
direct QAOA have 60/60. Because reconstructed distributions are truncated to
top 20 states whereas the monolithic distribution is complete, their EVS and
probability columns have different scopes and do not constitute a
matched full-distribution comparison. These data support a limited conclusion:
the coverage extension persists under these two simulated channels, while all
methods show distributional degradation under the stronger channel.

% Depth generalization ($p{=}1$--$3$). 
To verify that the extended-applicability result holds independently of
circuit depth, we sweep $p \in \{1,2,3\}$ on 40 dense instances
(Fig.~\ref{fig:e2e-depth}).
DC-QAOA's coverage is exactly 70\% at every depth, increasing
expressibility cannot overcome a partitioning infeasibility; the failure is
structural.
FrozenLGP ($B_f{=}3$) maintains 100\% coverage at all $p$; AR rises
from $0.977$ ($p{=}1$) to $0.988$ ($p{=}3$) following the standard QAOA
convergence trend.
The coverage independence from $p$ is theoretically expected: partitioning (freezing and, where needed, CPP) modifies the graph topology before any QAOA circuit is constructed, so the resulting sub-problems are valid regardless of circuit depth.

\section{End-to-End Effect of the Freeze Budget}
\label{subsec:e2e-budget}

The partition-level analysis identified $B_f \approx \kappa-(k-1)$ as the
coverage knee; the end-to-end question is whether quality survives as the
budget (and hence the amount of freezing) grows.
Fig.~\ref{fig:e2e-budget} answers affirmatively on the 30 hardest
instances (ER $p{=}0.8$ and BA $d{=}3$ at $n\in\{12,16,20\}$, where DC-QAOA achieves only 50\% coverage): approximation ratios are
flat across all budgets, 0.975, 0.968, 0.978, 0.973, 0.979 for
$B_f = 1$ through $5$, all pairwise CIs overlapping, while coverage is
100\% at every $B_f$.
What changes is the \emph{route} to coverage: at $B_f{=}1$, 37\% of
instances require the CPP fallback, declining to 27\% at $B_f{=}3$ and
13\% at $B_f{=}5$; the latter shift is accompanied by a steep increase in mean sub-circuit executions (Fig.~\ref{fig:e2e-budget}(b)).
Together with the partition-level threshold analysis, this supports a
simple operating rule: set $B_f \approx \kappa - (k-1)$ (2--3 for the
densities studied here); larger budgets buy no additional quality and
only inflate the quantum workload.

\section{CPP rescoring-rule ablation}
\label{app:cpp-ablation}

\begin{table*}[t]
\caption{\label{tab:cpp-ablation}%
CPP rescoring stress test on the eight CPP-triggered graphs among a
40-instance, four-family suite ($n\in\{10,12,14,16\}$; best of three paired
restarts). EVS and probabilities are conditional on the retained top-20
candidate distribution. All modes rank the same reconstructed candidates from
identical circuit outputs. ``Full objective'' ranks each retained parent
bitstring by its complete MaxCut objective and is the production default.}
\centering
\begin{tabular}{lrrrrr}
\hline
Rescoring rule & AR mean & AR median & EVS & $P_{\rm opt}$ & $P_{\ge.95}$ \\
\hline
None (MDR score only) & 0.987 & 0.992 & 0.955 & 0.294 & 0.624 \\
Additive & 0.987 & 0.992 & 0.952 & 0.300 & 0.555 \\
Multiplicative & 0.987 & 0.992 & 0.958 & 0.356 & 0.612 \\
Weighted & 0.992 & 1.000 & 0.963 & 0.380 & 0.669 \\
Full parent objective [default] & 0.992 & 1.000 & 0.978 & 0.519 & 0.839 \\
\hline
\end{tabular}
\end{table*}

The stress benchmark covers complete, dense ER,
7-regular, and BA $d=7$ graphs. It contains 600 runs (40 instances, five
rescoring modes, three restarts); CPP is exercised in 120 run-level records,
corresponding to eight graphs. Circuit signatures are identical across modes,
so the comparison isolates classical ranking.

On the CPP subset, omitted cross-partition edges account for mean 52.4\% and
maximum 69.2\% of total edge weight. CPP is therefore not a small correction. Directly
ranking retained bitstrings by the complete parent objective yields the highest
EVS (0.978) and optimum mass (0.519) in this sample, and ties the weighted rule
for best-candidate AR. Full-objective ranking is therefore the production default. Additive, multiplicative,
and weighted rules remain available only to reproduce the ablation. Because the
candidate list is already truncated before this ranking, the procedure does not
recover omitted quantum probability mass and does not make CPP exact. Coverage
comes from the width-feasible split; ranking affects only which retained
candidate is returned.

\section{Quantum Resource Footprint}
\label{subsec:e2e-resources}

\begin{figure*}[t]
    \centering
    \includegraphics[width=\textwidth]{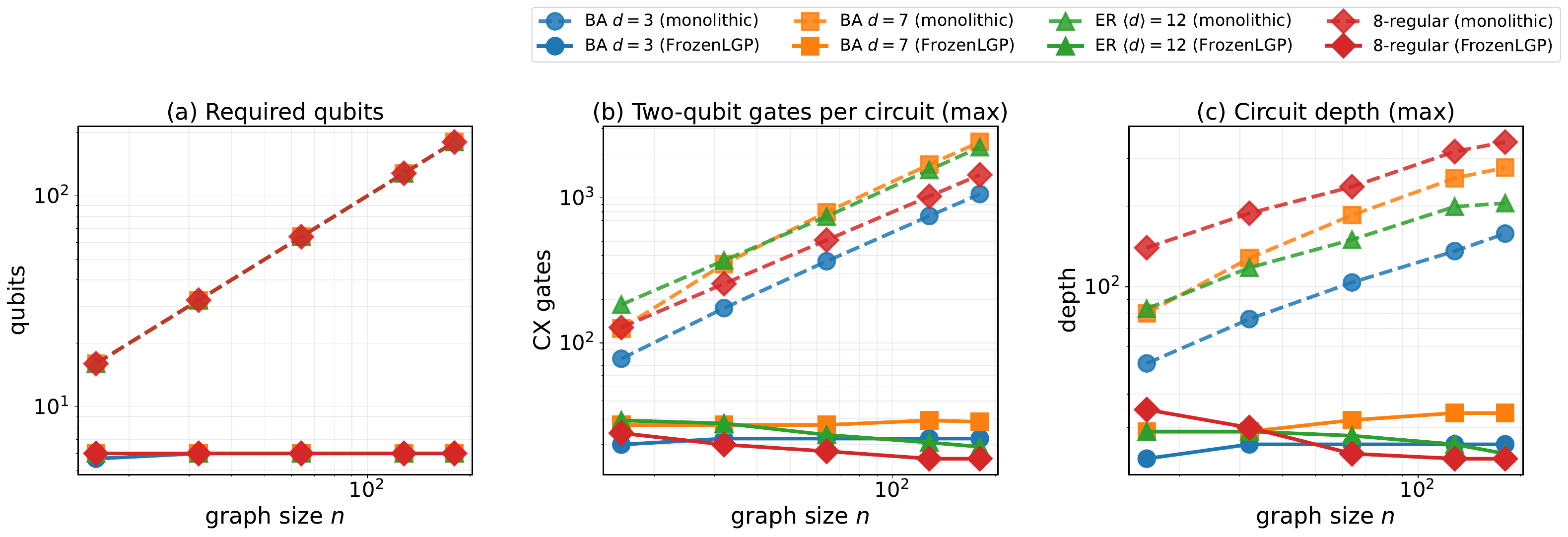}
    \caption{Transpiled circuit resources of monolithic QAOA (dashed)
    versus FrozenLGP sub-circuits (solid; maximum over leaves,
    $B_f{=}3$, $p{=}1$, CX basis, all-to-all connectivity, a lower
    bound favouring the monolithic baseline) for $n$ up to 180.
    (a)~Required qubits; (b)~two-qubit gates per executed circuit;
    (c)~circuit depth. These are per-circuit maxima; the figure does not
    include repeated assignments, optimiser evaluations, or shots and
    therefore is not a total-resource comparison.}
    \label{fig:e2e-resources}
\end{figure*}

\begin{table*}[t]
\caption{\label{tab:recursive-work}%
Recursive-work audit on 72 instances ($n\in\{12,16,20,24,32,50\}$, three
seeds, $k=6$, $B_f=3$). Structural leaves count distinct leaf graph structures.
Inflation is $W(\mathrm{root})$ from Eq.~\eqref{eq:recursive-work} divided by
that structural count. The path factor is the largest product of assignment
factors along any root-to-leaf path. The recorded run uses the exact partition
driver on subproblems with at most 20 vertices and the scalable driver above
20.}
\centering
\begin{tabular}{lrrrr}
\hline
Family & Instances & Median leaves & Inflation median & Inflation max / path max \\
\hline
BA $d=3$ & 18 & 17 & 1.0 & 1.0 / 1 \\
BA $d=7$ & 18 & 42 & 3.5 & 1,172,084.4 / 4,194,304 \\
Dense ER & 18 & 20 & 4.0 & 22.7 / 32 \\
7-regular & 18 & 37 & 3.0 & 4.0 / 4 \\
\hline
\end{tabular}
\end{table*}

Fig.~\ref{fig:e2e-resources} establishes a width/depth fact only. At
$n=180$ the transpiled monolithic circuit uses 180 qubits, about 1{,}800 CX
gates, and depth about 250 under all-to-all connectivity, whereas each
FrozenLGP leaf uses at most six qubits, about $22\pm8$ CX gates, and depth about
27. A small leaf can be easier to place on restricted-connectivity hardware,
but these per-circuit reductions do not imply a reduction in total executions
or total error exposure.

The telemetry counters make this distinction concrete. Under the shared caps in
Table~\ref{tab:qaoa2-resources}, FrozenLGP uses mean 65{,}431 shots and 0.759 million
two-qubit-gate--shots per instance, compared with 54{,}506 shots and 0.497
million gate--shots for QAOA\textsuperscript{2}. Under the original default
settings, FrozenLGP reaches 101{,}611 shots and 1.206 million gate--shots. These are aggregate experimental resources, not an inference from subcircuit counts alone.

The recursive-work audit exposes the worst recursive case. Equation
~\eqref{eq:recursive-work} reconciles exactly with explicit expansion on all 24
instances for which both were computed. Median assignment inflation ranges from
1 on separable BA $d=3$ graphs to four on dense ER, but one BA $d=7$ instance
has inflation $1.17\times10^6$ and a root-to-leaf assignment product
$2^{22}=4{,}194{,}304$ (Table~\ref{tab:recursive-work}). Thus the local
$2^{m_{\max}}$ bound cannot be advertised as a bound on the total algorithm.
Independent branches may be parallelised, but dependencies along a path remain;
hardware scheduling and wall-clock performance are not evaluated here.

% \section*{References}
\bibliographystyle{iopart-num-long}
\providecommand{\newblock}{}

\end{document}